\documentclass[reprint,
superscriptaddress,
 amsmath,amssymb,
 aps,
]{revtex4-2}

\usepackage{graphicx}
\usepackage{dcolumn}
\usepackage{bm}
\usepackage{xcolor}
\usepackage{gensymb}
\usepackage{adjustbox}
\usepackage{hyperref}
\usepackage[mathlines]{lineno}
\graphicspath{{images/}}

\begin{document}

\preprint{APS/123-QED}

\title{Forced Imbibition in Stratified Porous Media: \\Fluid Dynamics and \textcolor{black}{Breakthrough} Saturation}

\author{Nancy B. Lu}
\thanks{Denotes equal contribution.}
\affiliation{Department of Chemical and Biological Engineering, Princeton University, Princeton, NJ 08544}
\author{Daniel B. Amchin}
\thanks{Denotes equal contribution.}
\affiliation{Department of Chemical and Biological Engineering, Princeton University, Princeton, NJ 08544}
 \author{Sujit S. Datta}
 \thanks{To whom correspondence should be addressed:\\ssdatta@princeton.edu.}
\affiliation{Department of Chemical and Biological Engineering, Princeton University, Princeton, NJ 08544}

\date{\today}
 
\begin{abstract}
Imbibition, the displacement of a nonwetting fluid by a wetting fluid, plays a central role in diverse energy, environmental, and industrial processes. While this process is typically studied in homogeneous porous media with uniform permeabilities, in many cases, the media have multiple parallel strata of different permeabilities. How such stratification impacts the fluid dynamics of imbibition, as well as the \textcolor{black}{fluid saturation after the wetting fluid breaks through to the end of a given medium}, is poorly understood. We address this gap in knowledge by developing an analytical model of imbibition in a porous medium with two parallel strata, combined with a pore network model that explicitly describes fluid crossflow between the strata. By numerically solving these models, we examine the fluid dynamics and \textcolor{black}{fluid saturation left after breakthrough}. We find that the \textcolor{black}{breakthrough} saturation of nonwetting fluid is minimized when the imposed capillary number Ca is tuned to a value Ca$^*$ that depends on both the structure of the medium and the viscosity ratio between the two fluids. Our results thus provide quantitative guidelines for predicting and controlling flow in stratified porous media, with implications for water remediation, oil/gas recovery, and applications requiring moisture management in diverse materials. 
\end{abstract}

\maketitle


\section{Introduction}
\noindent Imbibition, the process by which a wetting fluid displaces a nonwetting fluid from a porous medium, plays crucial roles in our lives. It underlies key energy processes, such as oil/gas recovery \cite{morrow2001recovery, mattax1962imbibition, li2000characterization,nicolaides2015impact} and water management in fuel cells \cite{forner2016advanced}; environmental processes, such as geological CO$_2$ sequestration \cite{bennion2006supercritical,bennion2010drainage, hatiboglu2008pore,celia2015status}, groundwater aquifer remediation \cite{schaefer2000experimental,esposito2011remediation}, and moisture infiltration in soil and wood \cite{penvern2020,zhou2019,weisbrod2002imbibition,tesoro2007relative,hassanein2006investigation}; and diverse other applications including operation of chemical reactors \cite{attou1999two} and wicking in fabrics \cite{tang2015characterizing,tang2015characterizing2}, paper microfluidics 
\cite{castro2017characterizing}, building materials \cite{fourmentin2016,rahman2015recycled,takahashi1996acoustic}, coatings \cite{abdelouahab}, and diagnostic devices \cite{hong2015dynamics}. As a result, the physics of imbibition has been studied widely. \textcolor{black}{A common simplifying assumption is that the medium is homogeneous, with uniformly-disordered pores of a single mean size with no spatial correlations between them. For such homogeneous media, decades of previous work have demonstrated} how different invasion dynamics and flow patterns can arise depending on the pore size, solid wettability, surface roughness, as well as the fluid viscosity, interfacial tension, and flow boundary conditions \cite{lenormand1984role, chang2009experimental, hatiboglu2008pore, sun2016micro, hughes2000pore, lenormand1988numerical, lenormand1990liquids, sahimi1993flow, alava2004imbibition, stokes1986interfacial, weitz1987dynamic, hultmark2011influence,zhao2016wettability,tanino2018oil,odier2017forced}.

However, in many cases, porous media are not homogeneous. Instead, they can have parallel strata, characterized by different mean pore sizes, oriented along the direction of macroscopic fluid flow \cite{bear2013dynamics,galloway2012terrigenous,gasda2005upscaling,king2018microstructural}; these can further impact the interfacial dynamics in complex ways \cite{datta2013drainage,king2018microstructural,reyssat2009imbibition,zhou1997scaling,zapata1981theoretical,bear2013dynamics, galloway2012terrigenous,gasda2005upscaling}. Being able to predict and control imbibition in stratified porous media is therefore both fundamentally interesting and of critical practical importance. Some studies have indicated that heterogeneous invasion can arise during spontaneous imbibition from a bulk fluid reservoir \cite{ashraf2017spontaneous,ashraf2019capillary}, but did not investigate the distinct case of forced imbibition at a fixed flow rate. Other studies have provided tantalizing evidence that different heterogeneous invasion behaviors can arise during forced imbibition, but only probed a narrow range of flow rates \cite{cinar2004experimental, dong2005immiscible, dawe1992experimental}, only focused on bulk fluid saturation \cite{cinar2004experimental,dong1998characterization,zapata1981theoretical}, or explored small differences in pore size between strata \cite{dawe1992experimental,cinar2004experimental, ahmed1988experimental,zapata1981theoretical,yokoyama1981effects,debbabi2017viscous,Debbabi2017, ambekar2018interface}. A systematic study of these dynamics is lacking, and thus the physics of imbibition in stratified porous media remains poorly understood. 

Our previous work took a first step towards addressing these gaps in knowledge by experimentally visualizing the onset of forced imbibition in stratified porous media \cite{lu2020forced}. Our work explored a wide range of imposed volumetric flow rates $Q$ in media with parallel strata characterized by a large permeability ratio, arising from different mean pore sizes in the strata. We found that imbibition is spatially heterogeneous and exhibits an unexpected transition in fluid displacement behavior: for small $Q$, the wetting fluid preferentially invades the fine stratum of a smaller mean pore size, while for sufficiently large $Q$, the fluid instead preferentially invades the coarse stratum of a larger mean pore size. This transition\textcolor{black}{, which is not captured by models of imbibition that assume a homogeneous medium overall,} reflects the complex interplay between pore-scale capillary forces driving imbibition in the different strata, transverse viscous forces due to crossflow between strata, and longitudinal viscous forces due to macroscopic flow through the medium. However, while these results provide insight into the initial nature of imbibition at the inlet of the medium, they do not consider the subsequent dynamics as the fluid continues to move through the medium and ultimately breaks through at the outlet. 

These dynamics can play a critical role in many of the applications mentioned above. In particular, they determine two key quantities that are often used as metrics of performance: the breakthrough time $t_b$ and the \textcolor{black}{breakthrough} nonwetting fluid saturation $\textcolor{black}{S_{O}}$. For example, in oil/gas recovery \cite{king1999predicting, abrams1975influence} and groundwater remediation \cite{maji2008arsenic,neukum2009quantitative}, a wetting fluid is often used to displace and remove nonwetting fluid, such as hydrocarbons and non-aqueous contaminants. In these cases, the fluid properties and injection conditions must be chosen to most rapidly remove the largest quantity of nonwetting fluid, and therefore minimize both $t_b$ and $\textcolor{black}{S_{O}}$. However, due to the dearth of predictive models of imbibition, current approaches rely on trial and error. Hence, an improved understanding of the full dynamics of forced imbibition in stratified porous media is needed.

Here, we combine analytical theory with computational pore network modeling to examine these full dynamics. First, to gain intuition, we develop a theoretical model of imbibition in a porous medium with two parallel strata by considering the balance of capillary and viscous forces, neglecting crossflow between the strata. This balance is quantified by the global capillary number Ca $\equiv\mu_{w}(Q/A)/\gamma$ defined using macroscopic quantities, where $\mu_{w}$ is the wetting fluid dynamic shear viscosity, $A$ is the overall cross-section area of the medium, and $\gamma$ is the interfacial tension between the fluids. Next, to assess the influence of crossflow, we design a pore network model that explicitly treats this behavior. By numerically solving both models, we explore the full dynamics of imbibition; in particular, we investigate how the performance metrics $t_b$ and $\textcolor{black}{S_{O}}$ depend on Ca as well as the structure of the medium and the viscosity ratio of the two fluids. Our findings suggest a simple rule: to minimize both $t_b$ and $\textcolor{black}{S_{O}}$, as is often desirable, Ca should be tuned to a transition value Ca$^*$ that emerges from our theory, contrary to the conventional belief that Ca should be maximized. \textcolor{black}{Taken together, these results thus demonstrate that -- in addition other commonly-considered factors such as wettability and fluid properties -- stratification is another important factor that must also be considered when modeling imbibition in porous media. More broadly, our work provides} quantitative guidelines for predicting and controlling flow in stratified porous media, with implications for water remediation, oil/gas recovery, and applications requiring moisture management in diverse materials. 



\begin{figure}[htp!]
    \centering
    \includegraphics[width=0.75\columnwidth]{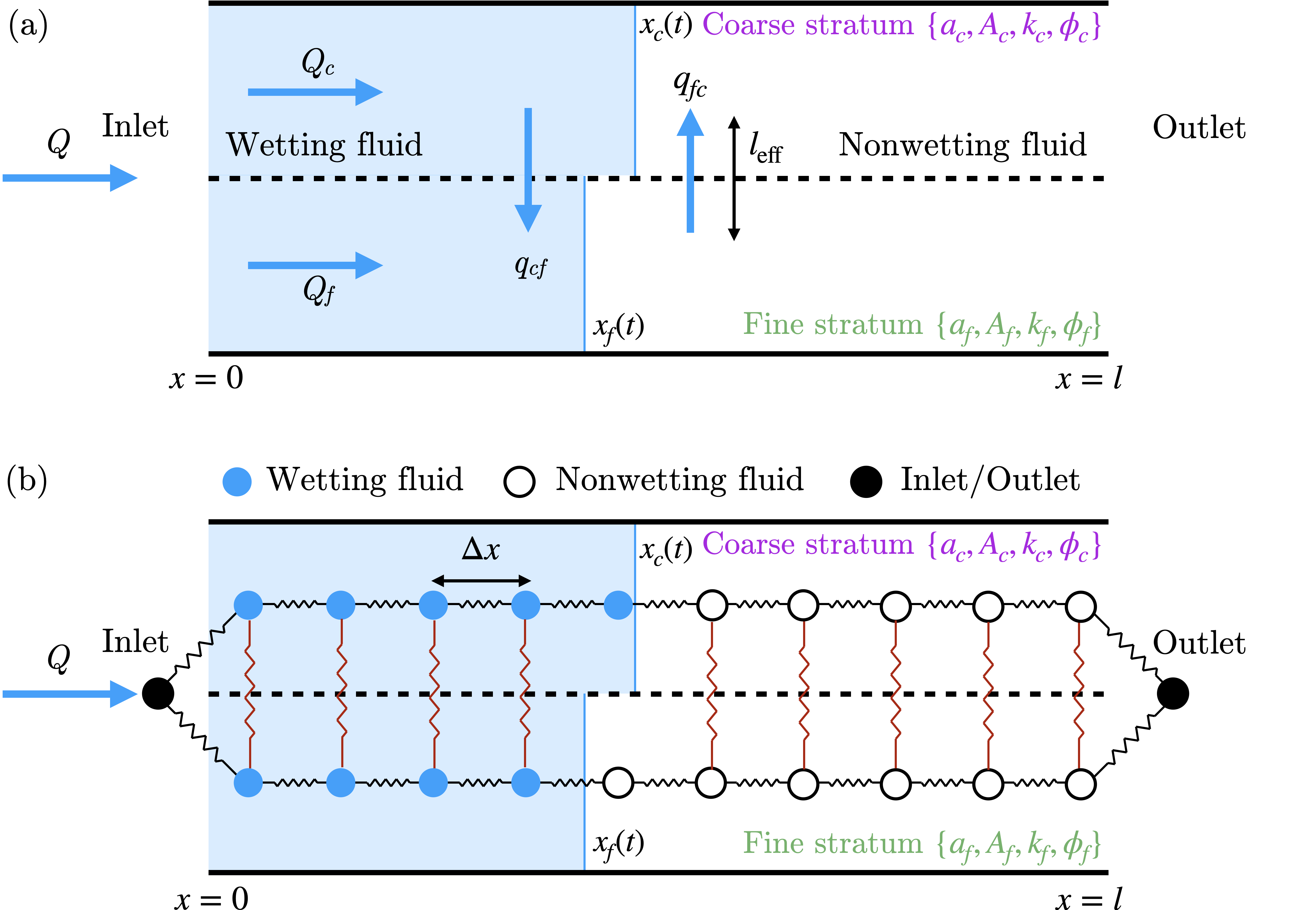}
    \caption{(a) Schematic of forced imbibition in a stratified medium with crossflow. Blue indicates wetting fluid and white indicates nonwetting fluid. The subscripts $\{c,f\}$ denote the coarse and fine stratum, respectively. $Q$ represents the total imposed volumetric flow rate. $Q_c$ and $Q_f$ represent the volumetric flow rate into the coarse and fine stratum, respectively. $q_{cf}$ and $q_{fc}$ represent the volumetric crossflow rate per length. $x_c(t)$ and $x_f(t)$ represent the position of the wetting/nonwetting fluid interface in each stratum. $\{a_i,A_i,k_i,\phi_i\}$ represent structural descriptors of each stratum $i$. (b) Schematic of pore network model. Each stratum is modeled as a series of nodes separated by a distance $\Delta x$ through which resistors flow. The black resistors represent longitudinal flow, while red resistors represent transverse crossflow. \textcolor{black}{Our theory and simulations do not rely on a specific pore morphology; however, \cite{lu2020forced} provides experimental images of an example stratified medium.}}
    \label{network_model}
\end{figure}

\section{Theoretical Model\\For Imbibition Dynamics}
\noindent As a first step toward modeling the complex flow behavior during imbibition, we consider a porous medium with parallel coarse and fine strata, characterized by uniform pore throat radii $a_i$, cross-sectional areas $A_i$, permeabilities $k_i$, and porosities $\phi_i$; hereafter $i\in\{c,f\}$ denotes the coarse and fine stratum, respectively. A schematic of this model is shown in Figure \ref{network_model}(a). Following our previous work
\cite{lu2020forced}, we analyze the distribution of pressures $p$ in each fluid and each stratum during imbibition. The pressure gradient in each stratum and fluid is described by Darcy's law $\frac{\partial p_{i}}{\partial x}=-\frac{\mu_{m} Q_{i}}{k_{i}A_{i}}$, where $m\in\{w,nw\}$ denotes the wetting and nonwetting fluids, respectively. Furthermore, at the interface between the fluids, the pressure in the wetting fluid is reduced by the capillary pressure $p_{c,i}\equiv2\gamma/a_{i}$ compared to the nonwetting fluid \cite{lenormand1983mechanisms,princen1969capillary,princen1969capillary2,princen1970capillary,mason1986meniscus}.

The flow rate can vary along the length of each stratum due to crossflow. Without loss of generality, we assume that the initial longitudinal position of the invading wetting/nonwetting fluid interface in the coarse stratum, $x_c$, is slightly ahead of the position of the invading interface in the fine stratum, $x_f$. Because the strata are in contact with each other over their entire length $\ell$, fluid can flow across them. We represent this crossflow by $q_{ij}$, the transverse volumetric flow rate from stratum $i$ to stratum $j$ per unit length along the medium; for simplicity, we consider that crossflow only occurs within a given fluid, i.e., there is no crossflow across different fluid phases. Mass conservation then implies that 
\begin{equation}
Q = Q_c + Q_f  
\label{eq:MassCons1}
\end{equation}
\begin{equation}
\mathrm{with}~Q_c=A_c \phi \frac{dx_c}{dt} \textcolor{black}{+} \int_0^{x_f} q_{cf}(x') ~dx'  
\label{eq:MassCons2}
\end{equation}
\begin{equation}
\mathrm{and}~Q_f=A_f \phi \frac{dx_f}{dt} \textcolor{black}{-} \int_0^{x_f} q_{cf}(x') ~dx',
\label{eq:MassCons3}
\end{equation}

\noindent where we have taken the porosity $\phi$ to be equal across the different strata for simplicity. 

Using Equations \ref{eq:MassCons1}, \ref{eq:MassCons2}, and \ref{eq:MassCons3}, we then use Darcy's Law in the coarse stratum from the inlet to the outlet to obtain an expression for the total pressure drop across the medium of length $\ell$:
\begin{equation}
\begin{split}
& p_{inlet} - p_{outlet} = \\ 
& \int_0^{x_f} \frac{\mu_{w}}{k_c} \left( \frac{Q_c - \int_0^x q_{cf}(x')~dx'}{A_c}\right) ~dx \\
& + \int_{x_f}^{x_c} \frac{\mu_{w}}{k_c} \left( \frac{Q_c - \int_0^{x_f} q_{cf}(x')~dx'}{A_c}\right) ~dx
\\ & + \int_{x_c}^{\ell} \frac{\mu_{nw}}{k_c} \left( \frac{Q_c - \int_0^{x_f} q_{cf}(x')~dx' + \int_{x_c}^{x} q_{fc}(x')~dx'}{A_c}\right) ~dx   
\\ & - p_{c,c}.
\end{split}
\label{eq:Darcy1}
\end{equation}
\noindent Repeating this procedure for the fine stratum results in another expression for the total pressure drop across the medium:
\begin{equation}
\begin{split}
& p_{inlet} - p_{outlet} = \\ 
& \int_0^{x_f} \frac{\mu_{w}}{k_f} \left( \frac{Q_f + \int_0^x q_{cf}(x')~dx'}{A_f}\right) ~dx 
\\ & +   \int_{x_f}^{x_c} \frac{\mu_{nw}}{k_f} \left( \frac{Q_f + \int_0^{x_f} q_{cf}(x')~dx'}{A_f}\right) ~dx
\\ & + \int_{x_c}^{\ell} \frac{\mu_{nw}}{k_f} \left( \frac{Q_f + \int_0^{x_f} q_{cf}(x')~dx' - \int_{x_c}^{x} q_{fc}(x')~dx'}{A_f}\right) ~dx  
\\ & - p_{c,f}.
\end{split}
\label{eq:Darcy2}
\end{equation}
\noindent Because the fluid pressure is equal across the strata at the inlet and outlet, Eq.~\eqref{eq:Darcy1} is subtracted from \eqref{eq:Darcy2}:
\begin{equation}
\begin{split}
& \mu_{w} \int_0^{x_f} \bigg[\left( \frac{Q_f + \int_0^x q_{cf}(x')~dx'}{k_f A_f}\right) 
\\ & - \left( \frac{Q_c - \int_0^x q_{cf}(x')~dx'}{k_c A_c}\right) \bigg] ~dx~ 
\\ & + \mu_{nw} \int_{x_f}^{x_c} \bigg[\left( \frac{Q_f + \int_0^{x_f} q_{cf}(x')~dx'}{k_f A_f}\right) 
\\ & -  \frac{\mu_{w}}{\mu_{nw}} \left( \frac{Q_c - \int_0^{x_f} q_{cf}(x')~dx'}{k_c A_c}\right)\bigg] ~dx~
\\ & + \mu_{nw} \int_{x_c}^{\ell} \bigg[\left( \frac{Q_f + \int_0^{x_f} q_{cf}(x')~dx' - \int_{x_c}^{x} q_{fc}(x')~dx'}{k_f A_f}\right) 
 \\ & - \left( \frac{Q_c- \int_0^{x_f} q_{cf}(x')~dx' + \int_{x_c}^{x} q_{fc}(x')~dx'}{k_c A_c}\right) \bigg] ~dx \\ &  -
 \left( p_{c,f} - p_{c,c} \right)= 0.
\end{split}
\label{eq:entirechannel}
\end{equation}\\

We introduce the \textit{ansatz} that a difference in pressures in the strata at a given longitudinal position $x$ drives a transverse crossflow between the strata: $q_{ij}(x) = \alpha_{ij} [p_i(x) - p_j(x)]$, where $\alpha_{ij}$ is a proportionality constant. To obtain the crossflow term, we integrate Darcy's law to some arbitrary distance $x$ and simply subtract the pressures obtained in each stratum at that location. This procedure yields the following integral equations for the crossflow in the wetting and nonwetting fluids, respectively:
\begin{equation}
\begin{split}
& \frac{q_{cf}}{\alpha_{cf} \mu_{w}} - \left( \frac{1}{k_c A_c} +  \frac{1}{k_f A_f} \right) \int_0^{x} \left( \int_0^\xi q_{cf}(x')~dx' \right)~d\xi 
\\ &  =  \left( \frac{Q_f}{k_f A_f} - \frac{Q_c}{k_c A_c} \right) x,
\end{split}
\label{eq: crossflowflow1}
\end{equation}

\begin{equation}
\begin{split}
& \frac{q_{fc}}{\alpha_{fc}} = 
\\& -\mu_{w} \int_0^{x_f} \Bigg[\left( \frac{Q_f + \int_0^\xi q_{cf}(x')~dx'}{k_f A_f}\right)  
\\ & - \left( \frac{Q_c - \int_0^\xi q_{ab}(x')~dx'}{k_c A_c}\right) \Bigg] ~d\xi~ 
\\ & -\mu_{nw} \int_{x_f}^{x_c} \Bigg[\left( \frac{Q_f + \int_0^{x_f} q_{cf}(x')~dx'}{k_f A_f}\right) 
\\ & -  \frac{\mu_{w}}{\mu_{nw}} \left( \frac{Q_c - \int_0^{x_f} q_{cf}(x')~dx'}{k_c A_c}\right)\Bigg] ~d\xi~
\\ & -\mu_{nw} \int_{x_c}^{x} \Bigg[\left( \frac{Q_f + \int_0^{x_f} q_{cf}(x')~dx' - \int_{x_c}^{\xi} q_{fc}(x')~dx'}{k_f A_f}\right) 
 \\ &- \left( \frac{Q_c- \int_0^{x_f} q_{cf}(x')~dx' + \int_{x_c}^{\xi} q_{fc}(x')~dx'}{k_c A_c}\right) \Bigg] ~d\xi  
\\ & +  \left( p_{c,f} - p_{c,c} \right).
\end{split}
\label{eq: crossflowflow2}
\end{equation}\\

\noindent Together, the six coupled integral equations Eqs.~\eqref{eq:MassCons1}, \eqref{eq:MassCons2}, \eqref{eq:MassCons3}, \eqref{eq:entirechannel}, \eqref{eq: crossflowflow1}, and \eqref{eq: crossflowflow2} provide a description of the full dynamics of imbibition in a stratified porous medium. However, in practice, analytically solving these equations is challenging. Therefore, in our previous work, we performed a linear stability analysis of these equations \cite{lu2020forced} to predict the flow behavior only at its incipient stages of imbibition. This analysis predicted two different types of behavior during the beginning of imbibition, which we confirmed using experiments: \textit{fine-preferential invasion}, where the wetting fluid invades the fine stratum faster, and \textit{coarse-preferential invasion}, where the wetting fluid invades the coarse stratum faster. The linear stability analysis does not capture any of the subsequent dynamics and therefore, does not enable predictions for the two key performance metrics $t_b$ and $\textcolor{black}{S_{O}}$. Hence, in this work, we turn to numerical simulations of Eqs.~\eqref{eq:MassCons1}, \eqref{eq:MassCons2}, \eqref{eq:MassCons3}, \eqref{eq:entirechannel}, \eqref{eq: crossflowflow1}, \eqref{eq: crossflowflow2} to examine these subsequent dynamics. 


\begin{table*}[htp!] 
\centering
\begin{adjustbox}{width=1\textwidth}
\begin{tabular}{|c|c|c|c|c|}
\hline
\textbf{Physical Parameter} & \textbf{Variable}  & \textbf{Experimental Range} & \textbf{Simulated Range} & \textbf{Reservoir Range} \\ \hline
Pore throat size ratio      & $a_c/a_f$          & 8.1                           & 1.4 to 14                & 1 to $\sim 10$ \cite{king2018microstructural}                            \\ \hline
Cross section area ratio    & $A_c/A_f$          & 0.2 to 4                    & 0.1 to 10                & $\sim0.05$ to $\sim20$ \cite{er2016characteristics}                           \\ \hline
Normalized medium length & $\ell/\sqrt{A}$    & 5 to 22                     & 0.7 to 70                & $\sim 10$ to $\sim 10^2$ \cite{SATTER2016195, yu2014simulation}                            \\ \hline
Fluid viscosity ratio             & $\mu_{nw}/\mu_{w}$ & 6.2                         & 0.1 to 10                & 0.6 to $ \sim 10^4$ \cite{arab2020water}                            \\ \hline
\rule{0pt}{10pt}Capillary number             & Ca & $\sim10^{-6}$ to $\sim10^{-4}$                         & $10^{-6}$ to  $10^{-3}$               & $\sim 10 ^{-12}$ to $\sim 10^{-5}$ \cite{abrams1975influence,yu2014simulation,SATTER2016195,arab2020water}                            \\ \hline
\end{tabular}
\end{adjustbox}
\caption{\textbf{Summary of relevant parameters.} Experimental range refers to our previous experiments  \cite{lu2020forced}, simulated range refers to the values tested in this work, and reservoir range refers to the values typically encountered in an oil reservoir.}
\label{experimental_values}
\end{table*}

A notable idealization of this theoretical model is that it treats the nonwetting/wetting fluid interface in each stratum as a flat front that completely displaces nonwetting fluid as it moves. However, interfacial instabilities are well-known to arise during flow, particularly at large values of the fluid viscosity ratio $\mu_{nw}/\mu_{w}$ \cite{lenormand1990liquids}. Under these conditions, the fluid interface can destabilize into discrete fingers, with different regions moving at different speeds. Moreover, discrete ganglia of nonwetting fluid can form and remain trapped behind the moving fingers. These additional behaviors are known to depend on a complex array of factors including flow rate, pore geometry, medium wettability, and pore size  \cite{lenormand1990liquids,iglauer2010x,iglauer2012comparison,georgiadis2013pore,wilkinson1984percolation, datta2014mobilization, krummel2013visualizing, blunt2017multiphase}; explicitly incorporating these additional physics into our model will be an important direction for future work. Nevertheless, the work presented here provides a useful first step that highlights the critical role played by stratification in influencing the dynamics of imbibition and consequently, the performance metrics $t_{b}$ and $\textcolor{black}{S_{O}}$.

\subsection{Imbibition Dynamics Without Crossflow}
\label{model_without_crossflow}

\noindent To gain intuition, we first simplify the model and neglect crossflow between the strata i.e., we set both $q_{fc}$ and $q_{cf}$ to be 0. Furthermore, to highlight the explicit dependence of the permeability on pore size and porosity, we apply the Kozeny-Carman relationship for permeability \cite{philipse93, sorbie1995extended}, $k_i=\phi^3 a_i^2 /[1.2(1-\phi)^2]$, hereafter. We numerically solve the simplified equations to determine the interface positions $x_c(t)$ and $x_f(t)$ at different imposed Ca. Using these simulations, we examine three key structural parameters: the ratio of the characteristic pore throat sizes of the two strata $a_c/a_f$, the ratio of the cross-section areas of the two strata $A_c/A_f$, and the nondimensional length of the overall medium $\ell/\sqrt{A}$, as well as the viscosity ratio $\mu_{nw}/\mu_{w}$. We use the experimental parameters from our previous work as the default values, but also sweep a broad range of these parameters, as summarized in Table \ref{experimental_values}. In each case, we plot the position of the nonwetting/wetting fluid interface in each stratum $i$ normalized by the overall length of the medium, $x_i/\ell$, as a function of time $t$ normalized by the characteristic invasion time $\tau\equiv\ell A \phi/Q$. This nondimensional time $t/\tau$ corresponds to the number of injected fluid pore volumes of the entire medium. The results obtained are shown in panel (a) of Figs. \ref{permeability_timepanel}, \ref{area_timepanel}, \ref{length_timepanel}, and \ref{viscosity_timepanel}.

In all cases, we observe the two different invasion behaviors consistent with our previous work: fine-preferential invasion at low Ca and coarse-preferential invasion at sufficiently large Ca. In between these regimes, \textit{transitional invasion} arises in which both strata are invaded at similar initial speeds---also as observed in our previous experiments \cite{lu2020forced}. Our previous linear stability analysis \cite{lu2020forced} predicted that the transition between these two behaviors occurs when Ca equals a transition capillary number:
\begin{equation}
\text{Ca}^*_0=\frac{2\mu_{w}/\mu_{nw}}{\ell(1/k_f-1/k_c)}\left(\frac{1}{a_f}-\frac{1}{a_c}\right). 
\label{castar0}
\end{equation}
This transition capillary number quantifies the value of Ca at which the difference in the characteristic viscous pressure drops across the overall strata, $\sim\frac{\mu_{nw} Q\ell}{A}\left(1/k_{f}-1/k_{c}\right)$, balances the difference in the characteristic capillary suction pressures in the strata, $\sim\gamma\left(1/a_{f}-1/a_{c}\right)$. The numerical simulations presented here show excellent agreement with this prediction in all cases, as shown by the blue lines in panel (a) of Figs. \ref{permeability_timepanel}, \ref{area_timepanel}, \ref{length_timepanel}, and \ref{viscosity_timepanel}. 

Our numerical simulations enable us to go beyond our previous prediction and reveal the full dynamics of imbibition over time, as detailed below. Furthermore, we use our simulations to assess how the performance metrics $t_b$, the wetting fluid breakthrough time, and $\textcolor{black}{S_{O}}$, the \textcolor{black}{breakthrough} nonwetting fluid saturation, depend on Ca as well as the structure of the medium and the fluid viscosity ratio, as shown in panel (a) of Figs. \ref{state_diagram_perm_ratio}, \ref{state_diagram_area_ratio}, \ref{state_diagram_length}, and \ref{state_diagram_viscosity}. We define $t_b$ as the moment at which the wetting fluid reaches the outlet in either stratum. To facilitate comparison across different simulations, we normalize $t_b$ by the characteristic invasion time $\tau$ evaluated at $\text{Ca}=4\times10^{-5}$, the median of the range of Ca explored in our simulations. At this moment of breakthrough, we also compute the \textcolor{black}{breakthrough} saturation as the volume fraction of nonwetting fluid remaining in the pore space of the adjacent, non-broken through stratum: $\textcolor{black}{S_{O}}=[A_cx_c(t_b)\phi+A_fx_f(t_b)\phi]/(A\ell\phi)$. As noted previously, our model assumes, as a simplifying approximation, that wetting fluid invasion into each stratum stably and completely displaces nonwetting fluid: all regions $x<x_{c}$ and $x<x_{f}$ are completely saturated with the wetting fluid, where $x_{c}$ and $x_{f}$ represent the positions of the flat fronts in the coarse and fine strata, respectively.

\subsection{Imbibition Dynamics With Crossflow}
\noindent Our simplified analysis neglecting crossflow between the strata provides useful intuition on how the dynamics of imbibition depend on the structure of the medium as well as the viscosities of the defending and invading fluids. Next, motivated by previous theoretical work \cite{zhou1997scaling, zapata1981theoretical}, we incorporate crossflow between strata $i$ and $j$ \textit{via} the position-dependent transverse flow rate per unit length along the medium $q_{ij}(x)=\alpha_{ij} [p_i(x) - p_j(x)]$, where the pressure difference $[p_i(x) - p_j(x)]$ across the interface between the strata drives crossflow and $\alpha_{ij}$ is a proportionality constant. In particular, the pressure difference drives a transverse volumetric flow rate of fluid across the strata, $Q_{ij}(x)$, which is defined within a differential width $dx$ centered at position $x$ along the medium: $Q_{ij}(x) =  q_{ij}(x)dx$, and therefore, $\alpha_{ij}=\frac{Q_{ij}(x)}{[p_i(x) - p_j(x)]dx}$. To estimate $\alpha_{ij}$, we make the Darcy-Brinkman-type assumption that the transverse flow is localized near the interface between strata, only extending over a characteristic length $\sim a_{i}$ or $\sim a_{j}$ away from the interface into stratum $i$ or $j$, respectively. Then, treating the transverse flow as proceeding between two porous media in series, $[p_i(x) - p_j(x)]=\frac{\mu Q_{ij}(x)}{hdx}\left(\frac{a_{i}}{k_{i}}+\frac{a_{j}}{k_{j}}\right)$ by Darcy's law; here $\mu$ is the dynamic shear viscosity of the fluid phase being considered and $h$ is the depth of the medium (i.e. into/out of the page in Fig. \ref{network_model}). Therefore,
\begin{equation}
    \alpha_{ij}=\frac{h}{\mu}\left(\frac{a_{i}}{k_{i}}+\frac{a_{j}}{k_{j}}\right)^{-1}.
    \label{alpha}
\end{equation}

\noindent The six coupled integral equations Eqs.~\eqref{eq:MassCons1}, \eqref{eq:MassCons2}, \eqref{eq:MassCons3}, \eqref{eq:entirechannel}, \eqref{eq: crossflowflow1}, \eqref{eq: crossflowflow2} with a nonzero $\alpha_{ij}$ given by Eq. \eqref{alpha} fully describe the evolution of the nonwetting/wetting fluid interface position in each of the two strata; however, analytically solving these equations is challenging. Hence, we instead develop a pore network model that numerically solves for the flow in a discretized representation of a stratified porous medium, as described in the next section.

\section{Pore Network Model}
\noindent Our network model is schematized in Fig. \ref{network_model}(b). Each stratum of the medium is represented by a one dimensional array of nodes connected by edges in series (black resistors). These two arrays lie adjacent to each other, and are connected using a transverse array of edges (red resistors) to incorporate crossflow---except for locations that have different fluids in the adjacent strata, at which crossflow is suspended (schematized by the missing red resistor in the middle of Fig. \ref{network_model}(b)). A single node introduces wetting fluid into both strata at the inlet with imposed volumetric flow rate $Q$, and both strata join together at another single node at the outlet, providing control of the overall boundary conditions. Each node represents a discrete position $x$ along the medium, with $\Delta x=\ell/50$ between successive nodes and the inlet and outlet corresponding to $x=0$ and $x=\ell$, respectively.

For prescribed $\{Q, A, \mu_{w}, \gamma\}$ -- which correspond to a given imposed value of Ca -- as well as prescribed values of $\{A_{c}, A_{f}, \mu_{nw}, a_{c}, a_{f}, \phi, \ell\}$, we solve for the fluid pressures in the nodes and flow rates in the edges at successive discrete time steps $\Delta t = 0.005\tau$, where $\tau\equiv\ell A \phi/Q$ is the characteristic invasion time of the entire medium. We do this using four equations: 
\renewcommand{\theenumi}{\roman{enumi}}
\begin{enumerate}

\item Conservation of mass and fluid incompressibility throughout dictates that all net flow $\sum Q_n$ into a given node $n$ must sum to zero.

\item For a given fluid in a given stratum, Darcy's law dictates that the flow rate through edge $e$ in stratum $i$ is $Q_{e,i}= (p_{n_{1}} - p_{n_{2}})\frac{ k_{e,i} A_{e,i}}{\mu_e \Delta x}$, where $p_{n_1}$ and $p_{n_2}$ are the pressures at the two nodes $n_{1}$ and $n_{2}$ connected by the edge, $k_{e,i}$ and $A_{e,i}$ describe the permeability and cross-section area of the stratum, and $\mu_e$ is the viscosity of the fluid type contained within the edge.

\item For the edge containing the immiscible fluid interface at a position $x_i$ in a given stratum $i$, we instead apply a modified form of Darcy's law that accounts for the fraction of the wetting fluid within the edge $f_{e,i}$ and the capillary pressure difference $p_{c,i}$ across the immiscible fluid interface: $Q_{e,i}= (p_{n_{1}} - p_{n_{2}}-p_{c,i})\frac{ k_{e,i} A_{e,i}}{\Delta x [\mu_{w}f_{e,i}+\mu_{nw}(1-f_{e,i})]}$.

\item Within a given fluid, we apply Darcy's law across the adjacent strata to describe crossflow: $ Q_{e,ij}= (p_{n_{1}} - p_{n_{2}}) \alpha_{ij} \Delta x $, where the edge $e$ now connects strata $i$ and $j$ and $\alpha_{ij}$ is given by Eq. \ref{alpha}. For the calculations given here, we choose $h=3$ mm and represent the overall cross-section of the medium as a square with $A=9\text{ mm}^2$ to facilitate comparison to our previous experiments \cite{lu2020forced}. 

\end{enumerate}


\noindent Together, the equations in i-iv constitute a system of linear equations that we solve numerically for the interface positions $x_c(t)$ and $x_f(t)$ until one of the interfaces breaks through to the outlet. To validate this network model, we first consider imbibition without crossflow ($\alpha_{ij}=0$). The solid lines in panel (a) of Figs. \ref{permeability_timepanel}, \ref{area_timepanel}, \ref{length_timepanel}, and \ref{viscosity_timepanel} show the results of the network model, while the open circles show the explicit numerical solutions of the coupled set of governing Eqs.~\eqref{eq:MassCons1}, \eqref{eq:MassCons2}, \eqref{eq:MassCons3}, \eqref{eq:entirechannel}, \eqref{eq: crossflowflow1}, \eqref{eq: crossflowflow2}. Both approaches yield identical results in all cases, confirming the fidelity of the network model. 

We next use the network model to examine the influence of crossflow on the dynamics of forced imbibition. Following the studies without crossflow described in Section \ref{model_without_crossflow}, we examine the influence of the three key structural parameters $a_c/a_f$, $A_c/A_f$, $\ell/\sqrt{A}$, as well as the viscosity ratio $\mu_{nw}/\mu_{w}$. The results obtained are summarized in panel (b) of Figs. \ref{permeability_timepanel}, \ref{area_timepanel}, \ref{length_timepanel}, and \ref{viscosity_timepanel}. 

In all cases, we observe similar dynamics to the case without crossflow. In particular, we again observe the two different invasion behaviors consistent with our previous work: fine-preferential invasion at low Ca and coarse-preferential invasion at sufficiently large Ca, with transitional invasion between the two. Our previous linear stability analysis \cite{lu2020forced} predicted that the transition between these two behaviors occurs when Ca equals a transition capillary number:  
\begin{equation}
\begin{split}
   & \text{Ca}^*=\text{Ca}^*_0 +
    \\ &  \frac{\mu_{w}}{\gamma\ell\left(1/k_{f}-1/k_{c}\right)}\left(\frac{1}{k_fA_f}+\frac{1}{k_cA_c}\right)\int_{0}^{\ell}\int_{0}^{\xi}q_{fc}(x)~dx~d\xi,
    \label{ca_Star}
    \end{split}
\end{equation}
where analysis of Eqs. \ref{eq:entirechannel} and \ref{eq: crossflowflow2} yields $q_{fc}(x)=\frac{\alpha_{fc}\left(p_{c,f} - p_{c,c}\right) \left(1-x/\ell\right)}{1 + \alpha_{fc} \left({\mu_{nw}}/{2}\right) \left( \frac{1}{k_c A_c} + \frac{1}{k_f A_f} \right) x (\ell-x)}$. We again find good agreement between the results of the network model and this theoretical prediction, as shown by the blue lines in panel (b) of Figs. \ref{permeability_timepanel}, \ref{area_timepanel}, \ref{length_timepanel}, and \ref{viscosity_timepanel}. The network model also enables us to go beyond this previous prediction and examine the full dynamics of imbibition over time. Furthermore, we again use the pore network model to investigate how the performance metrics $t_b$, the wetting fluid breakthrough time, and $\textcolor{black}{S_{O}}$, the \textcolor{black}{breakthrough} nonwetting fluid saturation, depend on Ca as well as the structure of the medium and the fluid viscosity ratio, as shown in panel (b) of Figs. \ref{state_diagram_perm_ratio}, \ref{state_diagram_area_ratio}, \ref{state_diagram_length}, and \ref{state_diagram_viscosity}.

\section{Results}
\subsection{Influence of Pore Throat Ratio}
\noindent The pore throat size contrast between strata is one important structural characteristic of a stratified medium. For example, in shales, the ratio of the pore throat sizes in a high permeability stratum compared to an adjacent low permeability stratum can be as high as 10 (Table \ref{experimental_values}). Hence, to explore the influence of this key structural descriptor, we vary the pore throat ratio $a_c/a_f$ from 1.4 to 14 by holding $a_c$ constant and changing $a_f$. Changing the pore throat ratio over this range has two impacts, because the permeability $k_{i}\sim a_{i}^2$ and the capillary pressure $p_{c,i}\sim a_{i}^{-1}$: it corresponds to a permeability ratio $k_c/k_f$ ranging from 2 to 200 and a capillary pressure ratio $p_{c,c}/p_{c,f}$ that concomitantly ranges from 0.7 to 0.07.

\begin{figure*}
    \centering
    \includegraphics[width=0.85\textwidth]{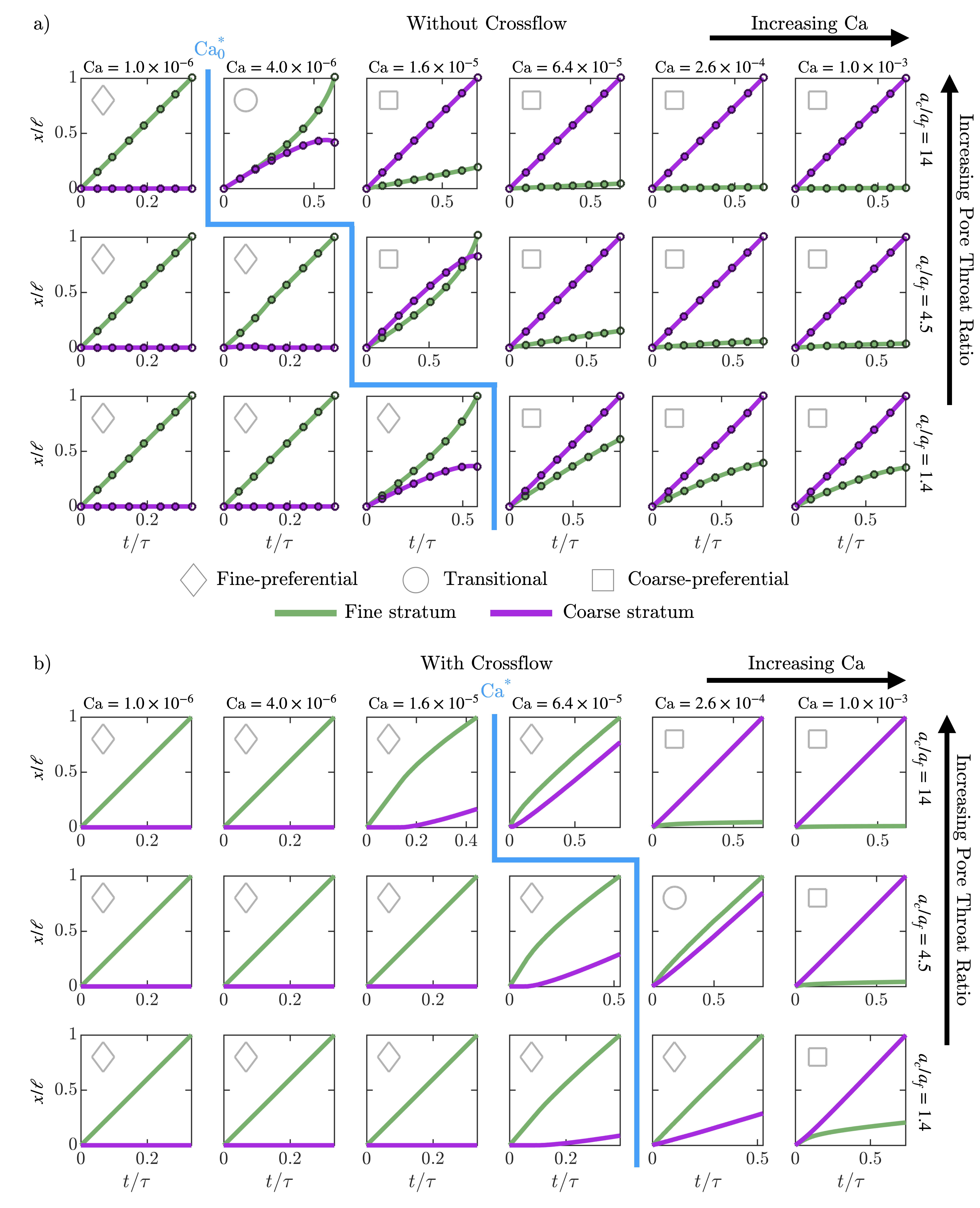}
    \caption{\textbf{Influence of pore throat ratio on imbibition dynamics (a) without crossflow and (b) with crossflow.} Plots show the normalized wetting/nonwetting fluid interface position as a function of normalized time. The open colored circles show the numerical solutions to Eqs.~\eqref{eq:MassCons1}, \eqref{eq:MassCons2}, \eqref{eq:MassCons3}, and \eqref{eq:entirechannel} to obtain the wetting/nonwetting fluid interface in each stratum $x_i(t)$. The connected solid lines show the results of the network model. Green and magenta indicate the fine and coarse stratum, respectively. The interface position $x$ is normalized by the length of the medium $\ell$ and time $t$ is normalized by the characteristic invasion time $\tau\equiv\ell A \phi/Q$. Diamonds indicate fine-preferential invasion. Circles indicate transitional invasion when the difference between the initial interface velocity in each stratum is $\leq 20\%$ the speed of the faster-moving interface. Squares indicate coarse-preferential invasion. The solid blue line indicates the value of (a) Ca$^*_0$ or (b) Ca$^*$ for each row. The normalized medium length is $\ell/\sqrt{A}=8.3$, the cross-section area ratio is $A_c/A_f=1$, and the viscosity ratio is $\mu_{nw}/\mu_{w}= 6.2$.}
    \label{permeability_timepanel}
\end{figure*}

\textit{Dynamics without crossflow:} First, we examine imbibition without crossflow, as shown in Fig. \ref{permeability_timepanel}(a). We initially focus on the smallest pore size ratio, $a_c/a_f=1.4$, as shown in the bottom row. At the lowest imposed Ca (first and second columns), the wetting fluid exclusively invades the fine stratum, as shown by the green points. We therefore classify this behavior as fine-preferential invasion, indicated by the diamond symbol in the top left corner. Under these conditions, capillarity dominates viscous effects; as a result, the wetting fluid imbibes faster into the fine stratum, for which the capillary suction $\sim a_{i}^{-1}$ is stronger. At a higher Ca $=1.6\times10^{-5}$ (third column), the coarse stratum is also invaded, albeit much slower. As invasion progresses, it speeds up in the fine stratum --- reflecting the absence of crossflow between the strata in this implementation of the model. In particular, the porous medium acts as a ``pore doublet" in which the growing viscous pressure drop limits continued invasion in the coarse stratum, while the enhanced capillary suction in the fine stratum continues to drive invasion into it --- redirecting flow from the former to the latter. Finally, at even higher Ca (fourth through sixth columns), the wetting fluid invades the coarse stratum increasingly faster than the fine stratum, as shown by the magenta points. We therefore classify this behavior as coarse-preferential invasion, indicated by the square symbols in the top left corners. Under these conditions, viscous effects increasingly dominate the influence of capillarity; as a result, the wetting fluid imbibes faster into the coarse stratum, for which the permeability $k_{i}\sim a_{i}^{2}$ is larger and thus the viscous pressure drop $\sim a_{i}^{-2}$ is lower. The onset of coarse invasion coincides with the predicted transition capillary number Ca$^*_0$ (Eq. \ref{castar0}), as shown by the blue line, confirming our previous prediction. 

\begin{figure*}
    \centering
    \includegraphics[width=0.85\textwidth]{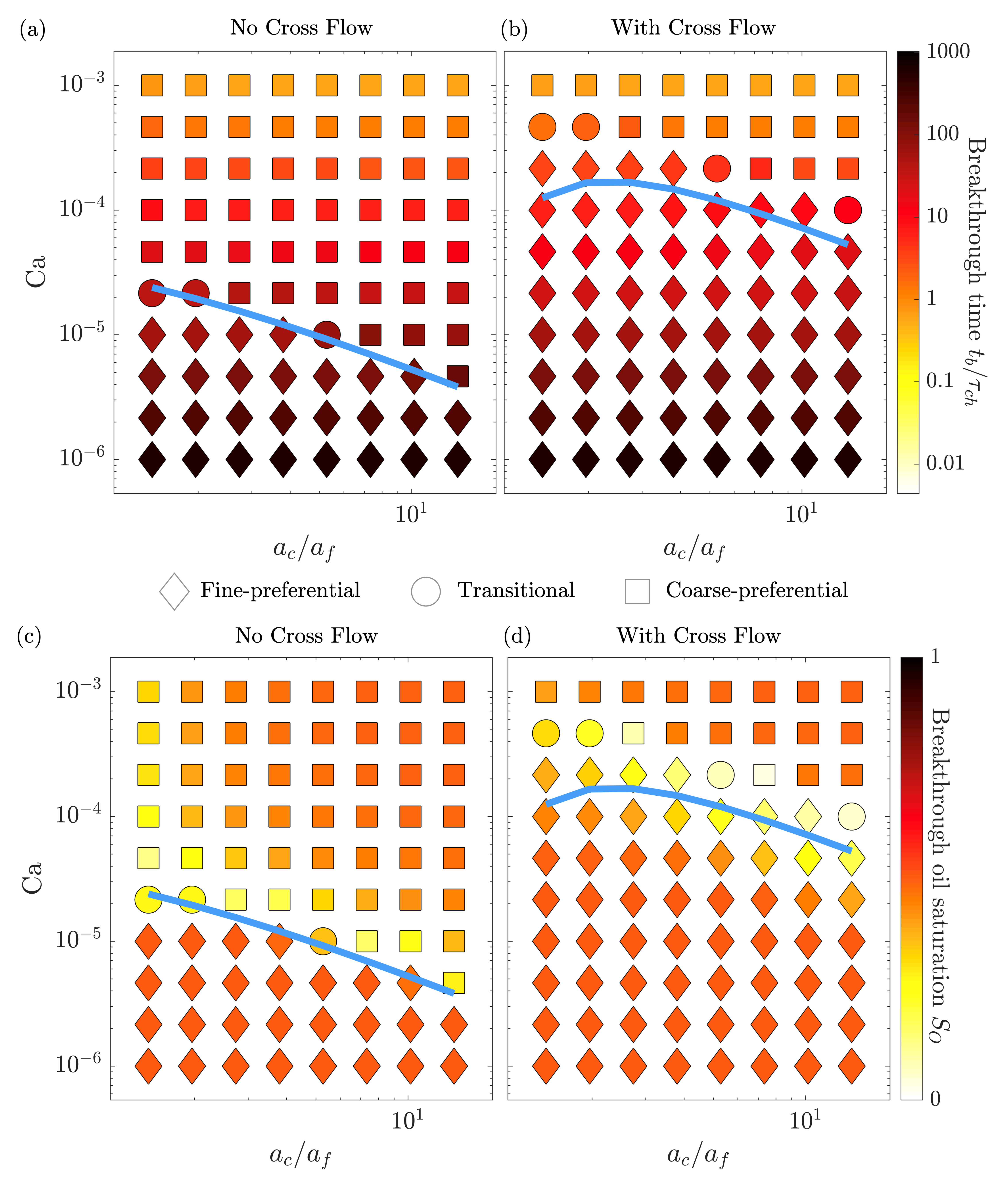}
    \caption{\textbf{Influence of pore throat ratio on (a),(b) breakthrough time and (c),(d) \textcolor{black}{breakthrough} nonwetting fluid saturation.} Plots show both metrics for imbibition (a),(c) without crossflow and (b),(d) with crossflow. (a),(b) Breakthrough time plotted as a function of imposed Ca and $a_c/a_f$. The color scale indicates the normalized breakthrough time $t_b/\tau_{ch}$ on a logarithmic scale, where $\tau_{ch}$ is fixed to be $\tau$ evaluated at $\text{Ca}=4.0\times10^{-5}$. (c),(d) \textcolor{black}{Breakthrough} nonwetting fluid saturation plotted as a function of imposed Ca and $a_c/a_f$. The color scale indicates the \textcolor{black}{breakthrough} saturation on a linear scale. For all panels, the normalized medium length is $\ell/\sqrt{A}=8.3$, the cross-section area ratio is $A_c/A_f=1$, and the viscosity ratio is $\mu_{nw}/\mu_{w}= 6.2$. The symbols are as described in the caption to Fig. \ref{permeability_timepanel}. The solid blue line indicates the value of Ca$^*_0$ for (a),(c) and of Ca$^*$ for (b),(d).}
    \label{state_diagram_perm_ratio}
\end{figure*}
We next investigate the influence of increasing pore throat ratio $a_c/a_f$. As exemplified by the middle and top rows of Fig. \ref{permeability_timepanel}(a), we find similar behavior as in the $a_c/a_f=1.4$ case, but with the onset of coarse invasion shifted to lower and lower Ca. Intriguingly, for the largest pore throat size ratio $a_{c}/a_{f}=14$, the transition between fine and coarse invasion is characterized by a regime in which both strata are initially invaded at similar speeds (Ca $=4.0\times10^{-6}$, second column): the difference between the initial invasion speeds in the different strata is $\leq 20\%$ the speed of the faster-moving interface. We therefore classify this behavior as transitional invasion, indicated by the circle symbol in the top left hand corner. Apparently similar behavior is observed near the transition (Ca $=1.6\times10^{-5}$) for the intermediate pore throat size ratio $a_{c}/a_{f}=4.5$, shown in the middle row and third column of Fig. \ref{permeability_timepanel}(a); however, in this case, the difference between the initial invasion speeds in the different strata is \textit{not} $\leq 20\%$ the speed of the faster-moving interface. We therefore still classify this behavior as coarse invasion, indicated by the square symbol in the top left hand corner. This example highlights that a classification based only on the incipient invasion behavior misses interesting subsequent dynamics during imbibition; indeed, as invasion progresses in this case, the interface speeds up in the fine stratum and recedes from the coarse stratum, again reflecting the ``pore doublet" dynamics described above. 

The shift in the transition between invasion behaviors is again consistent with our theoretical prediction for Ca$^*_0$, as indicated by the blue line. In particular, as given by Eq. \ref{castar0}, $\text{Ca}^*_0\sim\left(1/a_{f}-1/a_{c}\right)/\left(1/a_{f}^{2}-1/a_{c}^{2}\right)=1/\left(1/a_{f}+1/a_{c}\right)=a_{c}/\left(a_{c}/a_{f}+1\right)$, which scales as $\sim\left(a_{c}/a_{f}\right)^{-1}$ as pore throat size ratio increases: while capillary suction guides the initial invasion of the wetting fluid into the pores, viscous effects increasingly limit invasion into the lower-permeability fine stratum. Indeed, at the highest Ca (fourth through sixth columns), the wetting fluid invades the coarse stratum almost exclusively in the case of $a_{c}/a_{f}=14$ --- again reflecting coarse-preferential invasion. 

\textit{Dynamics with crossflow:} Incorporating crossflow into the model yields qualitatively similar invasion dynamics; compare Fig. \ref{permeability_timepanel}(b) to (a). In particular, in all cases, we observe a similar transition between fine- and coarse-preferential invasion around the Ca$^*$ predicted by Eq. \ref{ca_Star}, as shown by the blue line in Fig. \ref{permeability_timepanel}(b). Notably, $\text{Ca}^*>\text{Ca}^*_0$ in all cases: crossflow suppresses, but does not eliminate, the transition to coarse-preferential invasion. This effect arises from the influence of crossflow in forcing the nonwetting fluid from the fine stratum to the coarse stratum ahead of the fluid interface, reducing the overall viscous dissipation; therefore, one has to increase Ca even further for viscous effects to become comparable to capillarity. Additionally, unlike the case without crossflow, we do not observe ``pore doublet" dynamics in which the flow slows down in and even recedes from the coarse stratum: crossflow stabilizes these non-monotonic effects.


\textit{Breakthrough time:} The numerical simulations enable us to examine the breakthrough time $t_b$ both without and with crossflow incorporated, as shown in Figs. \ref{state_diagram_perm_ratio}(a) and (b), respectively. In each plot, we represent the normalized breakthrough time $t_{b}/\tau_{ch}$ by the color scale, the initial invasion behavior observed by the symbols shown, and the transition capillary number by the blue line. We observe similar results in both cases: as intuitively expected, the breakthrough time monotonically decreases with Ca---intriguingly, with no apparent dependence on $a_{c}/a_{f}$ or crossflow. Thus, processes that only seek to minimize breakthrough time should maximize the imposed Ca.

\textit{\textcolor{black}{Breakthrough} saturation:} We next examine the \textcolor{black}{breakthrough} nonwetting fluid saturation $\textcolor{black}{S_{O}}$ both without and with crossflow incorporated, as shown in Figs. \ref{state_diagram_perm_ratio}(c) and (d), respectively. We represent $\textcolor{black}{S_{O}}$ by the color scale, the initial invasion behavior observed by the symbols shown, and the transition capillary number again by the blue line. Contrary to the conventional belief that Ca should be maximized to minimize $\textcolor{black}{S_{O}}$, we find that in both cases, the \textcolor{black}{breakthrough} saturation exhibits a non-monotonic dependence on Ca: $\textcolor{black}{S_{O}}$ is minimized near the transition capillary number Ca$^*_0$ or Ca$^*$. Intuitively, this non-monotonic behavior reflects the improved displacement of nonwetting fluid during transitional invasion, in which \textit{both} strata are simultaneously invaded, instead of only one being preferentially invaded. Thus, processes that seek to minimize both breakthrough time and \textcolor{black}{breakthrough} saturation should \textit{not} maximize the imposed Ca. Instead, they should be tuned to slightly above the transition capillary number that describes the transition between invasion behaviors. The value of this transition capillary number decreases with increasing pore throat size ratio, highlighting the critical role played by the geometry of the medium in influencing performance during fluid imbibition.

\subsection{Influence of Cross-Section Area Ratio}
\noindent The cross-section area of each stratum is another key structural descriptor of a porous medium. The ratio of the cross-section areas between strata in a typical oil reservoir can range from 0.05 to 20 (Table \ref{experimental_values}). Hence, we vary the cross-section area ratio $A_c/A_f$ from 0.1 to 10 while keeping the total cross-section area $A\equiv A_f+A_c$ constant. 

\begin{figure*}
    \centering
    \includegraphics[width=0.85\textwidth]{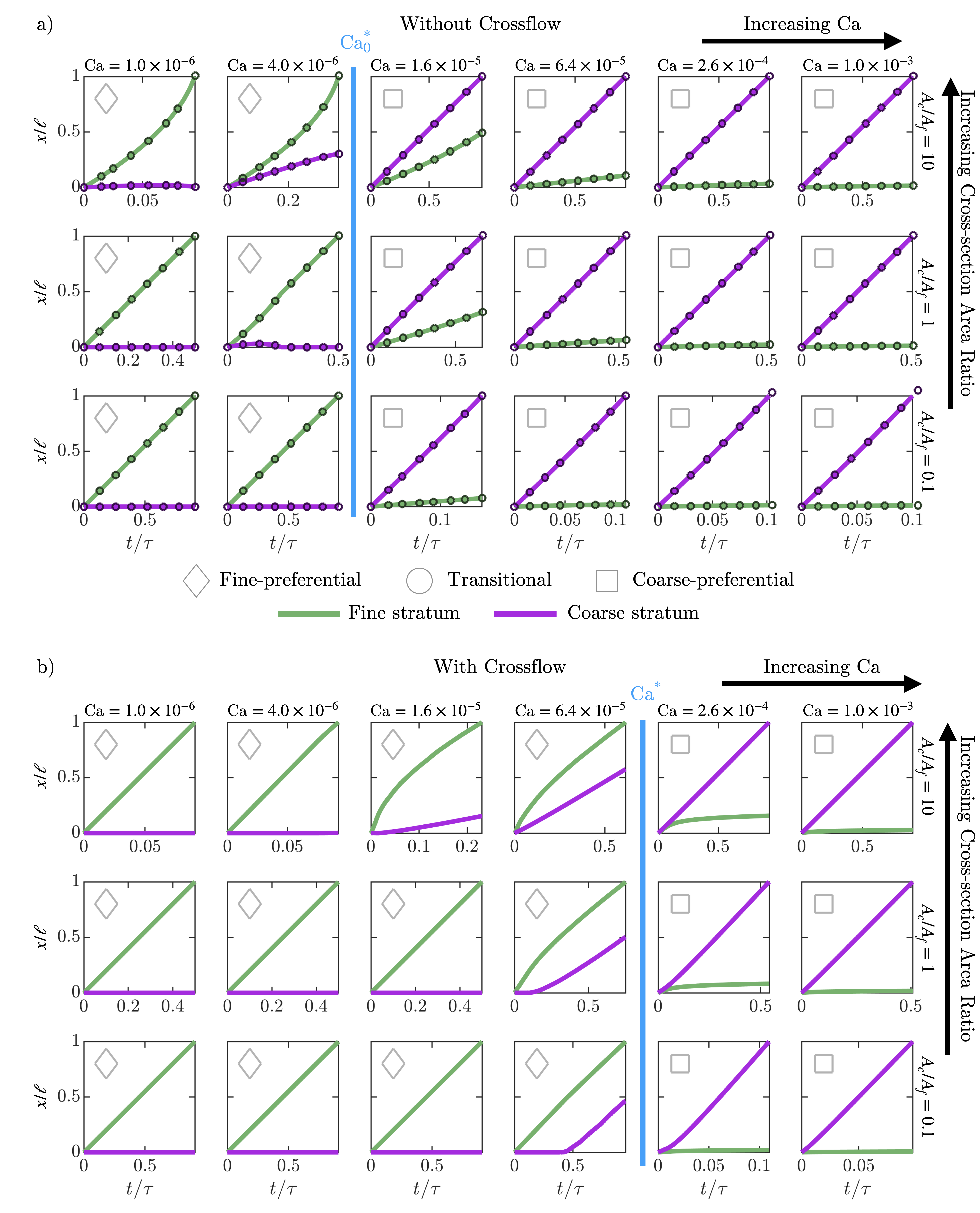}
    \caption{\textbf{Influence of cross-section area ratio on imbibition dynamics (a) without crossflow and (b) with crossflow.} Plots show the normalized wetting/nonwetting fluid interface position as a function of normalized time. The symbols, colors, and lines are as described in the caption to Fig. \ref{permeability_timepanel}. The normalized medium length is $\ell/\sqrt{A}=8.3$, the pore throat ratio is $a_c/a_f=8.1$, and the viscosity ratio is $\mu_{nw}/\mu_{w}= 6.2$.}
    \label{area_timepanel}
\end{figure*}

\textit{Dynamics without crossflow:} First, we examine imbibition without crossflow, as shown in Fig. \ref{area_timepanel}(a). We initially focus on the smallest cross-section area ratio, $A_c/A_f=0.1$, as shown in the bottom row. We again observe a transition between fine-preferential and coarse-preferential invasion driven by Ca, reflecting the dominance of pore-scale capillarity at low Ca and macroscopic viscous forces at high Ca. This transition again coincides with the predicted transition capillary number Ca$^*_0$, as shown by the blue line, confirming our previous prediction. We next investigate the influence of increasing cross-section area ratio $A_c/A_f$. As exemplified by the middle and top rows of Fig. \ref{area_timepanel}(a), we find similar behavior as in the $A_c/A_f=0.1$ case. This result is again consistent with our theoretical prediction for Ca$^*_0$, as indicated by the blue line. In particular, because the fluid pressure is equal across the strata at the inlet and outlet, $\text{Ca}^*_0$ (Eq. \ref{castar0}) is independent of $A_c/A_f$. Our simulations confirm this expectation: while we see slight differences in the full dynamics of imbibition across the different simulations, we observe no variation of $\text{Ca}^*_0$ with $A_c/A_f$.

\begin{figure*}
    \centering
    \includegraphics[width=0.85\textwidth]{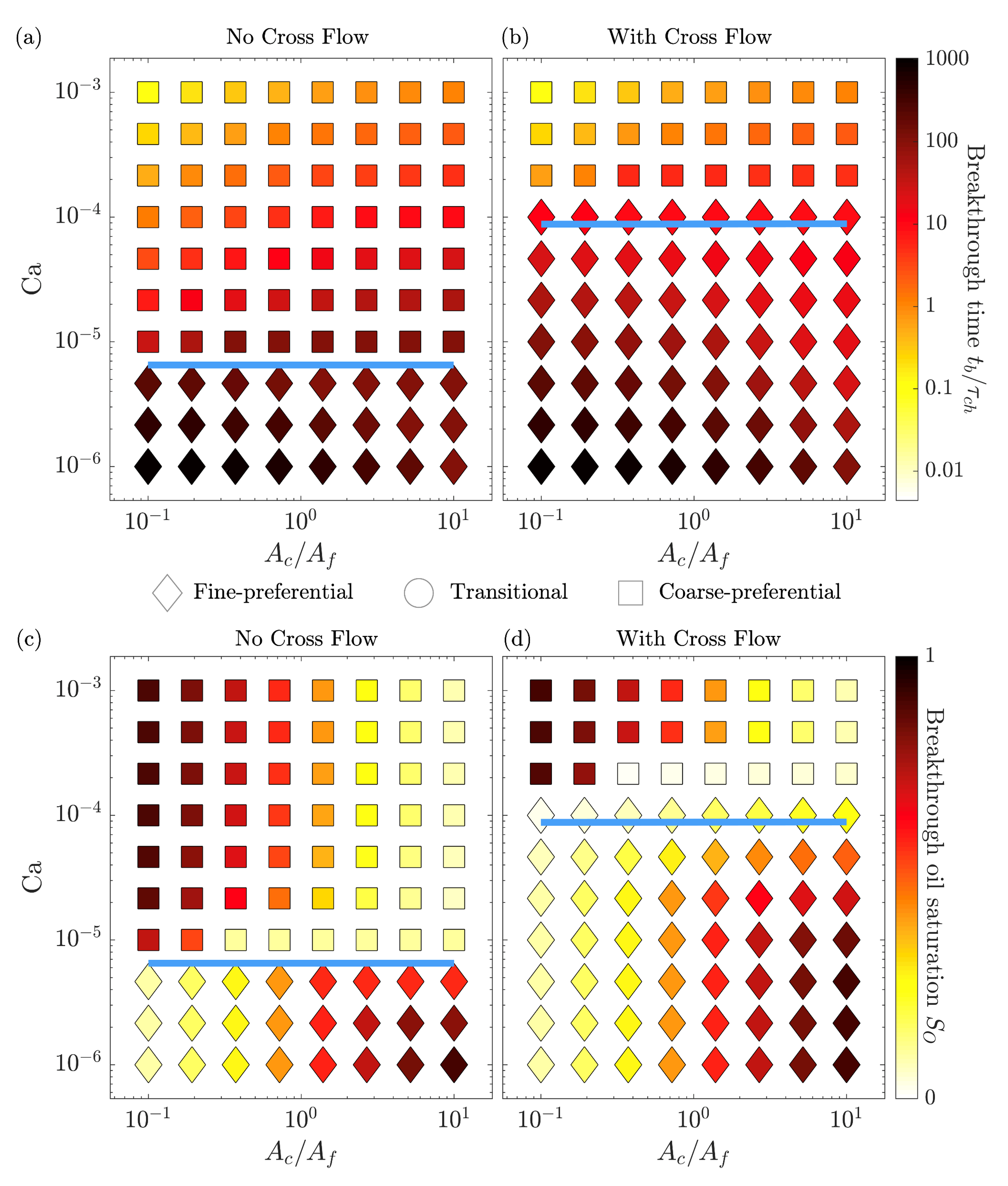}
    \caption{\textbf{Influence of cross-section area ratio on (a),(b) breakthrough time and (c),(d) \textcolor{black}{breakthrough} nonwetting fluid saturation.} Plots show both metrics for imbibition (a),(c) without crossflow and (b),(d) with crossflow. (a),(b) Breakthrough time plotted as a function of imposed Ca and $A_c/A_f$. (c),(d) \textcolor{black}{Breakthrough} nonwetting fluid saturation plotted as a function of imposed Ca and $A_c/A_f$. The color scales are described in the caption to Fig. \ref{state_diagram_perm_ratio}. For all panels, the normalized medium length is $\ell/\sqrt{A}=8.3$, the viscosity ratio is $\mu_{nw}/\mu_{w}= 6.2$, and the pore throat ratio is $a_c/a_f=8.1$. The symbols are as described in the caption to Fig. \ref{permeability_timepanel}. The solid blue line indicates the value of Ca$^*_0$ for (a),(c) and of Ca$^*$ for (b),(d).}
    \label{state_diagram_area_ratio}
\end{figure*}


\textit{Dynamics with crossflow:} Incorporating crossflow into the model yields qualitatively similar invasion dynamics; compare Fig. \ref{area_timepanel}(b) to (a). In particular, in all cases, we observe a similar transition between fine- and coarse-preferential invasion around the Ca$^*$ predicted by Eq. \ref{ca_Star}, as shown by the blue line in Fig. \ref{area_timepanel}(b). Again, $\text{Ca}^*>\text{Ca}^*_0$ in all cases: crossflow suppresses, but does not eliminate, the transition to coarse-preferential invasion. Furthermore, we again find that incorporating crossflow removes the non-monotonic ``pore doublet" dynamics. 

\textit{Breakthrough time:} The breakthrough times $t_b$ determined from the numerical simulations, both without and with crossflow incorporated, are shown in Figs. \ref{state_diagram_area_ratio}(a) and (b), respectively. We again observe similar results in both cases, but with the results below the transition Ca$^*_0$ shifted to higher values of Ca when crossflow is incorporated, reflecting the shift in the transition to Ca$^*$. As intuitively expected, for a given $A_{c}/A_{f}$, the breakthrough time monotonically decreases with Ca. Intriguingly, however, we observe different dependencies on $A_c/A_f$ for a given Ca below or above the transition value (Ca$^*_0$ or Ca$^*$); for Ca below the transition, $t_b$ decreases with increasing $A_c/A_f$, while conversely for Ca above the transition, $t_b$ increases with increasing $A_c/A_f$. These differing dynamics reflect the difference in invasion behaviors; in the case of fine-preferential invasion at low Ca, increasing $A_c/A_f$ reduces the volume of the fine stratum, and therefore the volume of fluid needed to be displaced, while for coarse-preferential invasion at high Ca, increasing $A_c/A_f$ increases the volume of the fluid that needs to be displaced from the coarse stratum. Thus, processes that only seek to minimize breakthrough time should maximize the imposed Ca and, if possible, minimize $A_c/A_f$.

\textit{\textcolor{black}{Breakthrough} saturation:} The \textcolor{black}{breakthrough} nonwetting fluid saturation $\textcolor{black}{S_{O}}$ determined from the numerical simulations, both without and with crossflow incorporated, are shown in Figs. \ref{state_diagram_area_ratio}(c) and (d), respectively. We again observe similar results in both cases, but with the results shifted to higher values of Ca when crossflow is incorporated, reflecting the shift in the transition from Ca$^*_0$ to Ca$^*$. Furthermore, we again observe different dependencies on $A_c/A_f$ below or above the transition value (Ca$^*_0$ or Ca$^*$), reflecting the difference in invasion behaviors. In the case of fine-preferential invasion at low Ca, increasing $A_c/A_f$ reduces the volume of the fine stratum from which nonwetting fluid is displaced, while in the case of coarse-preferential invasion at high Ca, increasing $A_c/A_f$ increases the volume of the coarse stratum from which nonwetting fluid is displaced. As a result, contrary to the conventional belief that Ca should be maximized to minimize $\textcolor{black}{S_{O}}$, we find that \textcolor{black}{breakthrough} saturation is minimized near the transition capillary number Ca$^*_0$ or Ca$^*$ --- again reflecting the improved displacement of nonwetting fluid from both strata during transitional invasion. Thus, processes that seek to minimize both breakthrough time and \textcolor{black}{breakthrough} saturation should \textit{not} maximize the imposed Ca. Instead, they should be tuned to be near the capillary number that describes the transition between invasion behaviors, independent of $A_c/A_f$.

\subsection{Influence of Medium Length}
\noindent The overall length $\ell$ of the medium is another key structural descriptor. For a fixed cross-section $A$, changing the length of a porous medium changes its overall aspect ratio; hence, to describe the length in a non-dimensional form, we normalize it by the characteristic transverse length $\sqrt{A}$. The normalized length of typical oil reservoirs can exceed 100 (Table \ref{experimental_values}). Hence, we vary the normalized length $\ell/\sqrt{A}$ from 0.67 to 67 while keeping the total cross-section area $A$ constant, which corresponds to $\ell$ varying from 0.2 to 20 cm. 

\begin{figure*}
    \centering
    \includegraphics[width=0.85\textwidth]{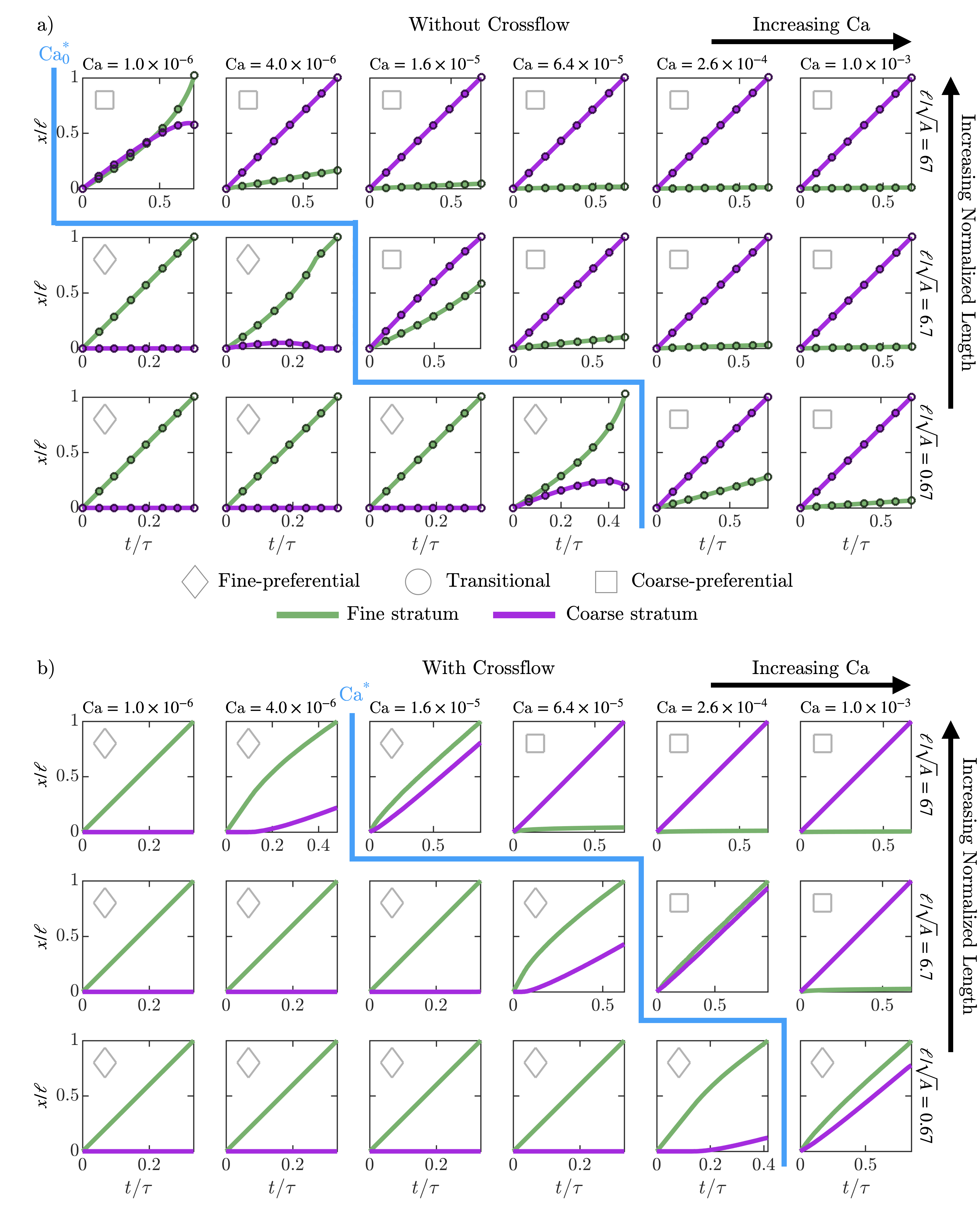}
    \caption{\textbf{Influence of medium length on imbibition dynamics (a) without crossflow and (b) with crossflow.} Plots show the normalized wetting/nonwetting fluid interface position as a function of normalized time. The symbols, colors, and lines are as described in the caption to Fig. \ref{permeability_timepanel}. The cross-section area ratio is $A_c/A_f=1$, the viscosity ratio is $\mu_{nw}/\mu_{w}= 6.2$, and the pore throat ratio is $a_c/a_f=8.1$.}
    \label{length_timepanel}
\end{figure*}
\begin{figure*}
    \centering
    \includegraphics[width=0.85\textwidth]{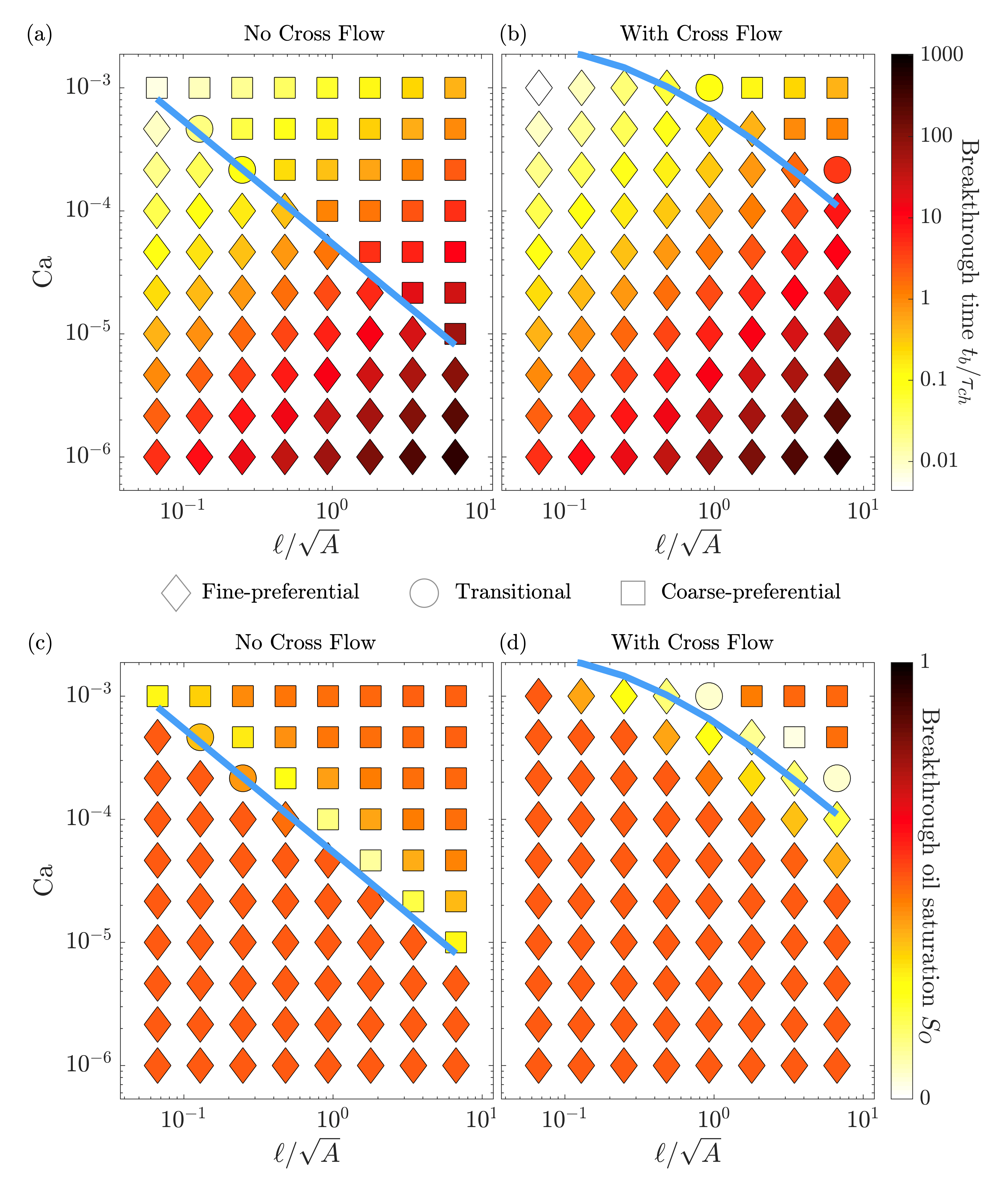}
    \caption{\textbf{Influence of medium length on (a),(b) breakthrough time and (c),(d) \textcolor{black}{breakthrough} nonwetting fluid saturation.} Plots show both metrics for imbibition (a),(c) without crossflow and (b),(d) with crossflow. (a),(b) Breakthrough time plotted as a function of imposed Ca and $\ell/\sqrt{A}$. (c),(d) \textcolor{black}{Breakthrough} nonwetting fluid saturation plotted as a function of imposed Ca and $\ell/\sqrt{A}$. The color scales are described in the caption to Fig. \ref{state_diagram_perm_ratio}. For all panels, the cross-section area ratio is $A_c/A_f=1$, the viscosity ratio is $\mu_{nw}/\mu_{w}= 6.2$, and the pore throat ratio is $a_c/a_f=8.1$. The symbols are as described in the caption to Fig. \ref{permeability_timepanel}. The solid blue line indicates the value of Ca$^*_0$ for (a),(c) and of Ca$^*$ for (b),(d).}
    \label{state_diagram_length}
\end{figure*}

\textit{Dynamics without crossflow:} For the case of imbibition wihtout crossflow as shown in Fig. \ref{length_timepanel}(a), we observe fine-preferential invasion at the lowest Ca, reflecting the dominance of pore-scale capillarity; by contrast, at sufficiently large Ca, we observe coarse-preferential invasion, reflecting the increasing influence of macroscopic viscous forces. Just above the transition (Ca $=1.0\times10^{-6}$) for the longest medium $\ell/\sqrt{A}=67$, shown in the top row and first column of Fig. \ref{length_timepanel}(a), we again observe apparent transitional invasion and long-time ``pore doublet" dynamics. This case is still classified as coarse invasion based on the initial invasion speeds---again highlighting that a classification based only on the incipient invasion behavior misses interesting subsequent dynamics during imbibition. In all cases, the transition between fine- and coarse-preferential invasion behaviors again coincides with the predicted transition capillary number Ca$^*_0$, as shown by the blue line, confirming our theoretical prediction (Eq. \ref{castar0}). In particular, because $\text{Ca}^*_0 \sim \ell^{-1}$, longer media --- for which the macroscopic viscous pressure drop is larger --- exhibit the transition to coarse-preferential invasion at lower Ca. 

\textit{Dynamics with crossflow:} Incorporating crossflow into the model again yields qualitatively similar invasion dynamics; compare Fig. \ref{length_timepanel}(b) to (a). In all cases, we observe a similar transition between fine- and coarse-preferential invasion around the Ca$^*$ predicted by Eq. \ref{ca_Star}, as shown by the blue line in Fig. \ref{length_timepanel}(b). Again, $\text{Ca}^*>\text{Ca}^*_0$ in all cases: crossflow suppresses, but does not eliminate, the transition to coarse-preferential invasion. Furthermore, we again find that incorporating crossflow removes the non-monotonic ``pore doublet" dynamics.


\textit{Breakthrough time:} The breakthrough times $t_b$ determined from the numerical simulations, both without and with crossflow incorporated, are shown in Figs. \ref{state_diagram_length}(a) and (b), respectively. We again observe similar results in both cases. For a given value of Ca, fastest breakthrough occurs at smaller $\ell/\sqrt{A}$ as intuitively expected -- the medium is shorter and thus will be invaded faster. Moreover, for a given $\ell/\sqrt{A}$, fastest breakthrough occurs at at higher Ca, again as intuitively expected. Thus, processes that only seek to minimize breakthrough time should maximize the imposed Ca and, if possible, minimize $\ell/\sqrt{A}$.

\textit{\textcolor{black}{Breakthrough} saturation:} The \textcolor{black}{breakthrough} nonwetting fluid saturation $\textcolor{black}{S_{O}}$ determined from the numerical simulations, both without and with crossflow incorporated, are shown in Figs. \ref{state_diagram_length}(c) and (d), respectively. Contrary to the conventional belief that Ca should be maximized to minimize $\textcolor{black}{S_{O}}$, we find that in both cases, the \textcolor{black}{breakthrough} saturation exhibits a non-monotonic dependence on Ca: $\textcolor{black}{S_{O}}$ is again minimized near the transition capillary number Ca$^*_0$ or Ca$^*$, once more reflecting the improved displacement of nonwetting fluid from both strata during transitional invasion. Thus, processes that seek to minimize both breakthrough time and \textcolor{black}{breakthrough} saturation should \textit{not} maximize the imposed Ca. Instead, they should be tuned to be near the capillary number that describes the transition between invasion behaviors. The value of this transition capillary number decreases with increasing normalized medium length, again highlighting the critical role played by the geometry of the medium in influencing performance during fluid imbibition.

\subsection{Influence of Viscosity Ratio}
\noindent Another key control parameter is the viscosity ratio between the nonwetting and wetting fluids, $\mu_{nw}/\mu_{w}$; for example, groundwater remediation and oil recovery operations often tune this parameter over a wide range for conformance control (Table \ref{experimental_values}). Hence, we probe $\mu_{nw}/\mu_{w}$ ranging from 0.1 to 10. Here, we accomplish this by holding $\mu_{w}$ constant and varying $\mu_{nw}$; the reverse case yields identical results (not shown). We reiterate that, while our results help to highlight the critical role played by stratification in influencing imbibition, they do not capture the full physics associated with interfacial instabilities that can arise during flow, particularly at large values of $\mu_{nw}/\mu_{w}$ \cite{lenormand1990liquids}. Incorporating these effects will be an important extension of our work.

\begin{figure*}
    \centering
    \includegraphics[width=0.85\textwidth]{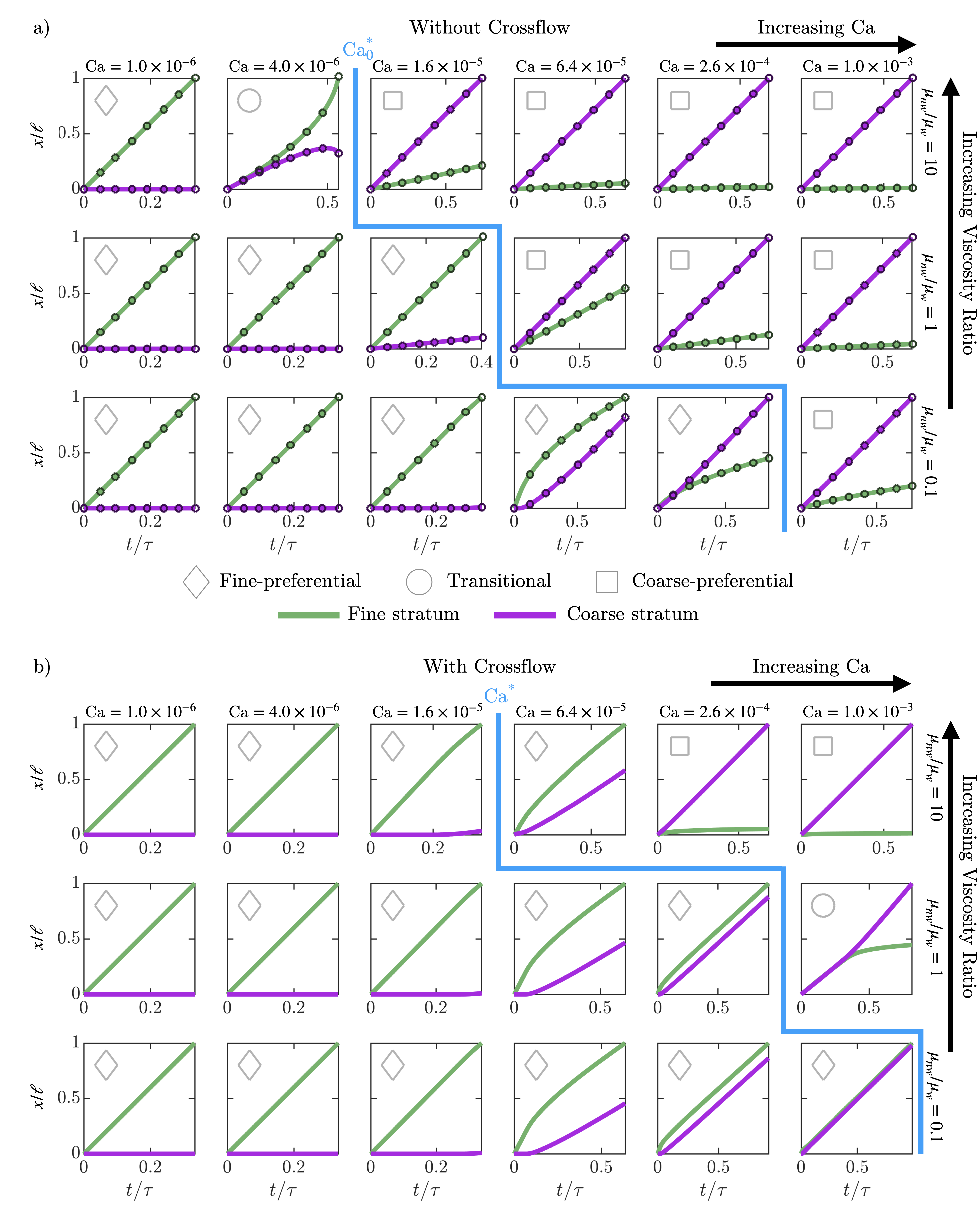}
    \caption{\textbf{Influence of viscosity ratio on imbibition dynamics (a) without crossflow and (b) with crossflow}. Plots show the normalized wetting/nonwetting fluid interface position as a function of normalized time. The symbols, colors, and lines are as described in the caption to Fig. \ref{permeability_timepanel}. The length of the medium is $\ell/\sqrt{A}=8.3$, the cross-section area ratio is $A_c/A_f=1$, and the pore throat ratio is $a_c/a_f=8.1$.}
    \label{viscosity_timepanel}
\end{figure*}
\begin{figure*}
    \centering
    \includegraphics[width=0.85\textwidth]{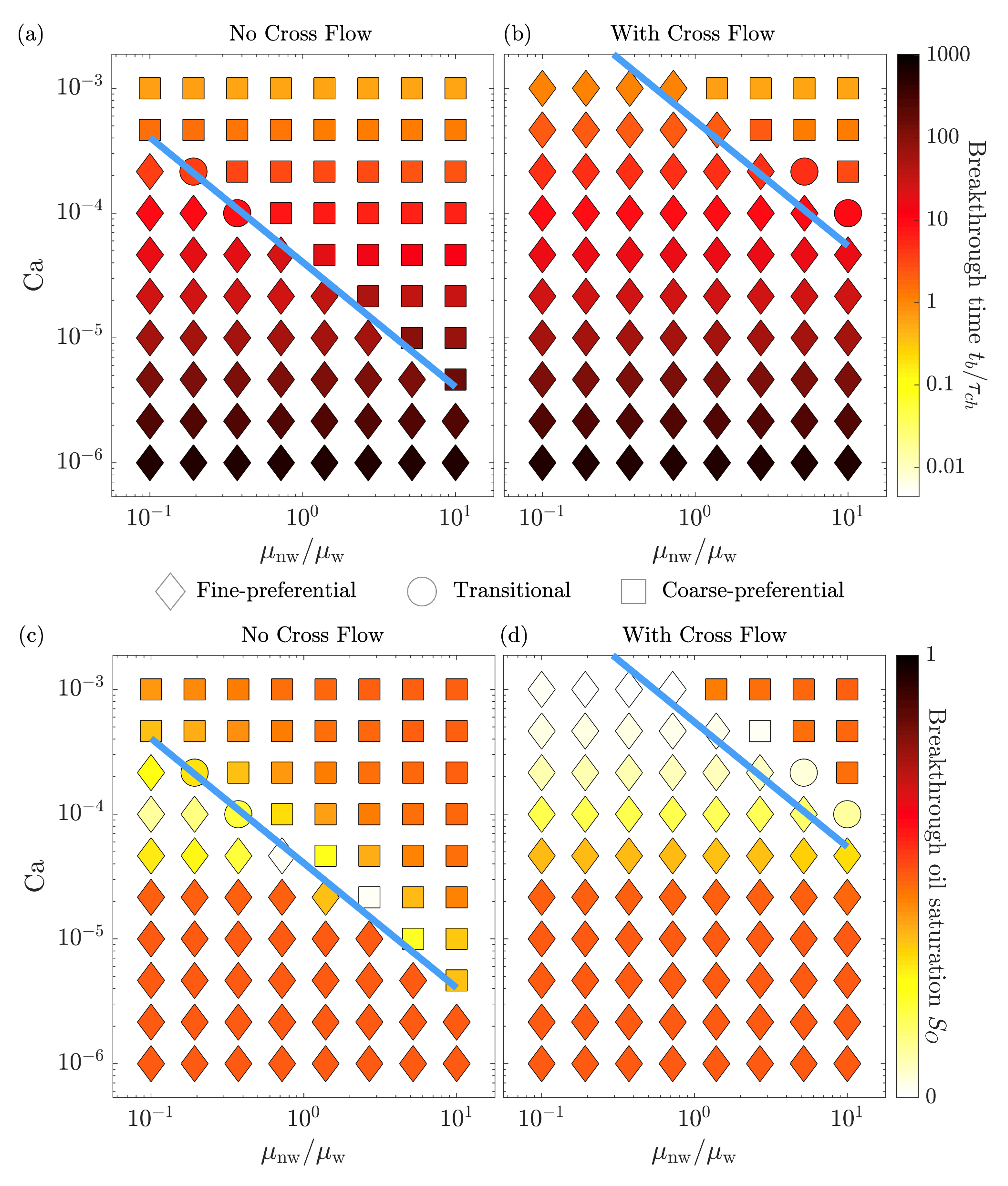}
    \caption{\textbf{Influence of viscosity ratio on (a),(b) breakthrough time and (c),(d) \textcolor{black}{breakthrough} nonwetting fluid saturation.} Plots show both metrics for imbibition (a),(c) without crossflow and (b),(d) with crossflow. (a),(b) Breakthrough time plotted as a function of imposed Ca and $\mu_{nw}/\mu_w$. (c),(d) \textcolor{black}{Breakthrough} nonwetting fluid saturation plotted as a function of imposed Ca and $\mu_{nw}/\mu_w$. The color scales are described in the caption to Fig. \ref{state_diagram_perm_ratio}. For all panels, the normalized medium length is $\ell/\sqrt{A}=8.3$, the cross-section area ratio is $A_c/A_f=1$, and the pore throat ratio is $a_c/a_f=8.1$. The symbols are as described in the caption to Fig. \ref{permeability_timepanel}. The solid blue line indicates the value of Ca$^*_0$ for (a),(c) and of Ca$^*$ for (b),(d).}
    \label{state_diagram_viscosity}
\end{figure*}

\textit{Dynamics without crossflow:} We first consider the dynamics of imbibition without crossflow, as shown in Fig. \ref{viscosity_timepanel}(a). At the lowest and highest Ca, we again observe fine- and coarse-preferential invasion, respectively, reflecting the competition between pore-scale capillarity and macroscopic viscous forces. Just below the transition for the smallest and largest viscosity ratios $\mu_{nw}/\mu_{w}=0.1$ and $10$, shown by the bottom and top rows, respectively, we again observe apparent transitional invasion and long-time ``pore doublet" dynamics. In all cases, the transition between fine- and coarse-preferential invasion behaviors again coincides with the predicted transition capillary number Ca$^*_0$, as shown by the blue line, confirming our theoretical prediction (Eq. \ref{castar0}). In particular, because $\text{Ca}^*_0 \sim \left(\mu_{nw}/\mu_{w}\right)^{-1}$, increasing the fluid viscosity ratio shifts the transition to coarse-preferential invasion at lower Ca: while capillary suction guides the initial invasion of the wetting fluid into the pores, viscous forces exerted by the invading fluid increasingly limit invasion into the lower-permeability fine stratum. 

\textit{Dynamics with crossflow:} Incorporating crossflow into the model again yields qualitatively similar invasion dynamics; compare Fig. \ref{viscosity_timepanel}(b) to (a). In all cases, we observe a similar transition between fine- and coarse-preferential invasion around the Ca$^*$ predicted by Eq. \ref{ca_Star}, as shown by the blue line in Fig. \ref{viscosity_timepanel}(b). Again, $\text{Ca}^*>\text{Ca}^*_0$ in all cases: crossflow suppresses, but does not eliminate, the transition to coarse-preferential invasion. Furthermore, we again find that incorporating crossflow removes the non-monotonic ``pore doublet" dynamics.

\textit{Breakthrough time:} The breakthrough times $t_b$ determined from the numerical simulations, both without and with crossflow incorporated, are shown in Figs. \ref{state_diagram_viscosity}(a) and (b), respectively. We again observe similar results in both cases: $t_b$ monotonically decreases with Ca, with no strong dependence on $\mu_{nw}/\mu_{w}$ or crossflow. Thus, within the assumptions of our model, processes that only seek to minimize breakthrough time should maximize the imposed Ca.

\textit{\textcolor{black}{Breakthrough} saturation:} The \textcolor{black}{breakthrough} nonwetting fluid saturation $\textcolor{black}{S_{O}}$ determined from the numerical simulations, both without and with crossflow incorporated, are shown in Figs. \ref{state_diagram_viscosity}(c) and (d), respectively. Contrary to the conventional belief that Ca should be maximized to minimize $\textcolor{black}{S_{O}}$, we find that in both cases, the \textcolor{black}{breakthrough} saturation exhibits a non-monotonic dependence on Ca: $\textcolor{black}{S_{O}}$ is again minimized near the transition capillary number Ca$^*_0$ or Ca$^*$, again reflecting the improved displacement of nonwetting fluid from both strata during transitional invasion. Additionally, for a fixed $\mu_{nw}/\mu_{w}$, $\textcolor{black}{S_{O}}$ decreases with increasing Ca up to the transition --- reflecting the broad range of capillary numbers over which increasingly transitional invasion arises. Thus, within the assumptions of our model, processes that seek to minimize both breakthrough time and \textcolor{black}{breakthrough} saturation should \textit{not} maximize the imposed Ca. Instead, they should be tuned to be near the capillary number that describes the transition between invasion behaviors, whose value decreases with increasing viscosity ratio,  highlighting the key role played by fluid formulation in influencing performance during imbibition.

\begin{figure*}
    \centering
    \includegraphics[width=0.85\textwidth]{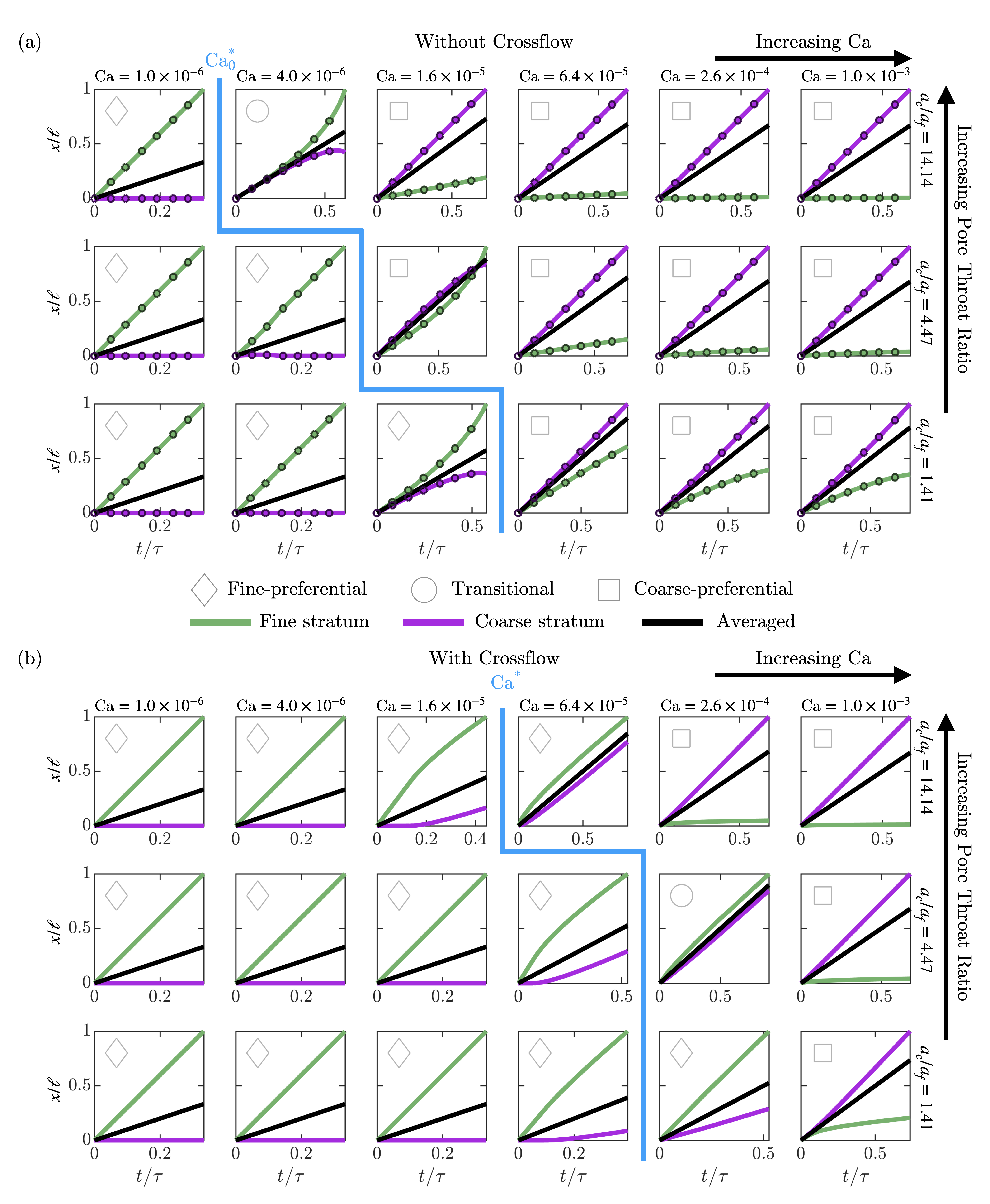}
    \caption{\textbf{Influence of averaging permeability between strata on imbibition dynamics (a) without crossflow and (b) with crossflow.} Plots show the normalized wetting/nonwetting fluid interface position as a function of normalized time. The open colored circles show the numerical solutions to Eqs.~\eqref{eq:MassCons1}, \eqref{eq:MassCons2}, \eqref{eq:MassCons3}, and \eqref{eq:entirechannel} to obtain the wetting/nonwetting fluid interface in each stratum $x_i(t)$. The connected solid lines show the results of the network model. Green and magenta indicate the fine and coarse stratum, respectively. The black line indicates a homogeneous medium with a permeability that is the area-weighted average of the two strata (i.e. $k_{\rm{avg}}=(A_fk_f+A_ck_c)/A)$ .The interface position $x$ is normalized by the length of the medium $\ell$ and time $t$ is normalized by the characteristic invasion time $\tau\equiv\ell A \phi/Q$. Diamonds indicate fine-preferential invasion. Circles indicate transitional invasion when the difference between the initial interface velocity in each stratum is $\leq 20\%$ the speed of the faster-moving interface. Squares indicate coarse-preferential invasion. The solid blue line indicates the value of (a) Ca$^*_0$ or (b) Ca$^*$ for each row. The normalized medium length is $\ell/\sqrt{A}=8.3$, the cross-section area ratio is $A_c/A_f=1$, and the viscosity ratio is $\mu_{nw}/\mu_{w}= 6.2$.}
    \label{averaged_permeability}
\end{figure*}

\subsection{Comparison to an averaged model}
\noindent In practice, imbibition dynamics are often modeled by treating stratified porous media as being non-stratified, with homogeneous, strata-averaged properties -- thus treating imbibition as being homogeneous across different strata. The work presented in this paper has revealed how such averaging can yield erroneous predictions. \textcolor{black}{As a final illustration of this point, we compare the predictions of our model explicitly incorporating heterogeneous invasion dynamics for media of different pore throat size ratios (Fig. \ref{permeability_timepanel}) to the predictions of a stratum-averaged model, in which the medium has a single effective permeability given by the area-weighted permeabilities of the separate strata: $k_{\rm{avg}}=(A_fk_f+A_ck_c)/(A_f+A_c)$. Because of the heterogeneous flow arising from stratification, in all cases, fluid breakthrough is considerably faster in the stratified medium than one would expect for a homogeneous medium --- as shown in Fig. \ref{averaged_permeability}. Compare, for example, the black curves (imbibition in a homogeneous medium) to the green and magenta curves (imbibition in the fine and coarse strata, respectively) in the top row of Fig. \ref{averaged_permeability}b. At the lowest Ca, fluid barely enters the coarse stratum; as a result, it breaks through the fine stratum rapidly, whereas at the same time, it would only have invaded $\sim40\%$ of a homogeneous medium. Conversely, at the highest Ca, fluid barely enters the fine stratum, and breaks through the coarse stratum rapidly, whereas it would have only invaded $\sim70\%$ of a homogeneous medium. Moreover, in both cases, the bypassing of either the coarse or fine stratum leaves behind $\sim50\%$ of the defending nonwetting fluid---a considerable amount that is completely missed by the averaged model, which does not capture any of the heterogeneous invasion behaviors that arise in stratified media, and therefore only approximates the flow behavior near the transitional invasion regime.}

\section{Discussion and Conclusion}
\noindent Our theoretical analysis and numerical simulations of the full dynamics of forced imbibition in a stratified medium quantify two distinct heterogeneous invasion behaviors, consistent with our and others' previous experimental results \cite{lu2020forced,cinar2004experimental, dong2005immiscible, dawe1992experimental,dong1998characterization,zapata1981theoretical,ahmed1988experimental,yokoyama1981effects,debbabi2017viscous,Debbabi2017, zapata1981theoretical, ambekar2018interface}: at small imposed capillary number Ca, the wetting fluid preferentially invades the fine stratum, while at large Ca, the wetting fluid instead preferentially invades the coarse stratum. We find that the transition between these behaviors is well described by the transition capillary number Ca$^*_0$ (Eq. \ref{castar0}) or Ca$^*$ (Eq. \ref{ca_Star}) -- describing imbibition without and with crossflow between the strata, respectively -- as predicted by our previous linear stability analysis \cite{lu2020forced}. Furthermore, our results suggest that crossflow between adjacent strata shifts the invasion dynamics to higher values of Ca, reflecting the shift in the transition from Ca$^*_0$ to Ca$^*$. Our numerical simulations enable us to examine how two key performance metrics -- the wetting fluid breakthrough time $t_b$ and the nonwetting fluid \textcolor{black}{breakthrough} saturation $\textcolor{black}{S_{O}}$ -- depend on injection conditions as well as the structure of the medium. In a homogeneous, averaged model, both quantities are minimized, as is often desirable, by simply maximizing the imposed Ca. Our work makes a starkly different prediction: to minimize both breakthrough time and \textcolor{black}{breakthrough} nonwetting fluid saturation, practitioners should tune Ca to near the transition capillary number, thereby preventing one stratum from being bypassed. Thus, we anticipate that this work will help guide imbibition processes in energy, environmental, and industrial applications that seek to control both $t_b$ and $\textcolor{black}{S_{O}}$. Two prominent examples include pump-and-treat remediation of contaminated groundwater aquifers \cite{bear2013dynamics,rabideau1994two} and enhanced oil/gas recovery \cite{king2018microstructural,ingham2010geomaterials,tanino2012capillary,bear2013dynamics,galloway2012terrigenous}; in both cases, injected wetting fluids are used to displace nonwetting fluid from heterogeneous subsurface media as quickly and effectively as possible. Other examples include controlling moisture wicking in building materials \cite{fourmentin2016,rahman2015recycled,takahashi1996acoustic}, layered microporous fabrics \cite{elnashar2005volume, adams1991permeability, ahn1991simultaneous}, and paper microfluidics \cite{castro2017characterizing, martinez2008three, conn2014visualizing,jang2015facile},

Together, this work highlights how stratification fundamentally alters the dynamics of forced imbibition in a porous medium. It represents a first step toward capturing all the physics underlying these complex dynamics, and necessarily involves some simplifying assumptions and approximations. For example, we treat the nonwetting/wetting fluid interface in each stratum as being sharp; incorporating the influence of pore-scale heterogeneities and interfacial instabilities will be an important extension of our work. Moreover, our model describes crossflow across strata within a given fluid using a Darcy's law model of transverse flow localized near the interface between the strata; more detailed studies of this crossflow, and incorporating crossflow with different fluids into the model, will be a useful direction for future work. Finally, while our work focuses on a porous medium with two adjacent strata as a model system, the results motivate further studies of flow in media with even more strata.

\begin{acknowledgments}
\noindent It is a pleasure to acknowledge Amir Pahlavan and Howard Stone, as well as Gary Hunter, Hubert King, and Jeremy Brandman from ExxonMobil Corporate Strategic Research, for stimulating discussions. This work was supported by ExxonMobil through its membership in the Princeton E-filliates Partnership of the Andlinger Center for Energy and the Environment and in part by the Grand Challenges Initiative of the High Meadows Environmental Institute, and in part by funding from the Princeton Center for Complex Materials, a Materials Research Science and Engineering Center supported by NSF grant DMR-2011750. N.B.L. was also supported in part by the Mary and Randall Hack Graduate Award of the High Meadows Environmental Institute.

\textit{Author contributions --- }N.B.L. and S.S.D. designed the study; N.B.L. numerically solved the theoretical model; D.B.A. designed and performed simulations of the pore network model; N.B.L. and S.S.D. analyzed the data; and S.S.D. designed and supervised the overall project. All authors discussed the results and wrote the manuscript.

\end{acknowledgments}

\end{document}